\documentclass[twocolumn,caps,showpacs]{revtex4}

\usepackage{graphicx,color}
\usepackage{latexsym}
\usepackage{amsmath,amssymb}        
\usepackage[draft=false]{hyperref}
\usepackage{mathrsfs}
\DeclareSymbolFontAlphabet{\mathrsfs}{rsfs}
\DeclareMathAlphabet{\mathcal}{OMS}{cmsy}{m}{n}

\begin{document}


\title{Behavior of luminous matter in the head-on encounter of two ultralight BEC dark matter halos}


\author{F. S. Guzm\'an}
\affiliation{Instituto de F\'{\i}sica y Matem\'{a}ticas, Universidad
              Michoacana de San Nicol\'as de Hidalgo. Edificio C-3, Cd.
              Universitaria, 58040 Morelia, Michoac\'{a}n,
              M\'{e}xico.}

\author{J. A. Gonz\'alez}
\affiliation{Instituto de F\'{\i}sica y Matem\'{a}ticas, Universidad
              Michoacana de San Nicol\'as de Hidalgo. Edificio C-3, Cd.
              Universitaria, 58040 Morelia, Michoac\'{a}n,
              M\'{e}xico.}

\author{J. P. Cruz}
\affiliation{Instituto de F\'{\i}sica y Matem\'{a}ticas, Universidad
              Michoacana de San Nicol\'as de Hidalgo. Edificio C-3, Cd.
              Universitaria, 58040 Morelia, Michoac\'{a}n,
              M\'{e}xico.}


\date{\today}


\begin{abstract}
Within the context of ultralight BEC dark matter, we analyze the head-on encounters of two structures. These structures are made of a BEC component, which is a ground state equilibrium solution of the Gross-Pitaevskii-Poisson system, together with a component of luminous matter. The evolution of the Condensate dark matter is carried out by solving the time dependent GPP equations, whereas the luminous matter is modeled with particles interacting gravitationally on top of the BEC dark matter halos. We track the evolution of frontal encounters for various values of the collision velocity and analyze the regime of high velocity regime showing solitonic behavior of the BEC halos and that of slow velocities producing a single final structure. We measure the relative velocity of the dark matter with respect to the luminous matter after the encounters in the solitonic case and track the evolution of luminous matter in the case of merger.
\end{abstract}



\maketitle

\section{Introduction}
\label{sec:introduction}

The Bose Einstein Condensate (BEC) or wave dark matter model has shown an important development, specially after the structure formation simulations showing some of the properties of the dark matter structures under this model \cite{WaveDM0,WaveDM}, and the potential it has shown in the solution of the cusp-core problem \cite{HarkoCC,Marsh,MatosCC,GuzmanLora2015GR} to mention a few interesting properties. This model basically assumes dark matter is a Bose-Einstein condensate made of spinless ultralight bosons  whose general properties and brief history can be found in \cite{MatosReview}.

At local scales, the system is ruled by the Schr\"odinger-Poisson (SP) system of equations, which turns into the Gross-Pitaevskii-Poisson (GPP) system of equations as soon as the wave function is interpreted as the descriptor of the mean-field approach of a Bose condensate \cite{GP}, submitted to a gravitational trap sourced by its own gravitational field. In this way, it is said that the evolution of this type of dark matter is ruled by the GPP system. Many aspects of the BEC dark matter, including the evolution of fluctuations have been deeply analyzed in stationary scenarios (see e.g. \cite{Chavanis,Chavanis2}).

In terms of the simulations in \cite{WaveDM}, the resulting high density configurations, made of standing waves are soliton solutions of the GPP system. Prior to the results in \cite{WaveDM}, the solitonic solutions of the GPP system were found to show solitonic behavior \cite{DaleChoi,BernalGuzman2006b}, that is, in a head-on collision, two configurations trespass each other and recover their initial shape and momentum after the encounter \cite{Solitons}. This is a rather non-trivial result, considering the gravitational potential is coupled to the GP equation in a non-trivial manner. In \cite{DaleChoi,BernalGuzman2006b} the conditions in terms of the total energy of the system were found to have solitonic behavior and form interference patterns during a collision, whereas dust-like dark matter under similar conditions did not behave in the same way \cite{GonzalezGuzman2011}. Moreover, the interference patterns eventually should imprint observable behavior to luminous matter.

If BEC dark matter is to show a particular behavior, some fingerprints on luminous matter are to be predicted. The effects of BEC dark matter local structures on luminous matter has been explored mainly related to galactic rotation curves (see e.g. \cite{RCs}), however the model has been tested barely on dynamical situations. Recently some interesting consequences of relative behavior between ultralight axion-dark matter and baryons show a possible impact on the formation rate of early stars and the heating of interstellar medium, based on a cosmological approach \cite{Marsh2015}. On the other hand, at local scales, in \cite{Paredes2015} an approximate analysis of the behavior of luminous matter with respect to BEC dark matter in collision of halos was developed, focusing specifically on the relative displacement between luminous and dark matter in the interaction of structures after an encounter and applied the results to the Abell 3827 cluster observed off-sets. 

The luminous matter so far has not been coupled to BEC dark matter in previous analyses. It has been considered to be a polytrope, for instance in \cite{MatosZeus}, and using a Miyamoto-Nagai disk model is evolved on top of a time independent BEC dark matter halo and a somehow unjustified hydrodynamical pressure is assumed; in this paper instead, we evolve the gravitational potential due to the BEC dark matter. In other cases, luminous matter is considered test particles traveling on the gravitational potential due to the BEC dark matter or it is modeled as a fluid ruled by the hydro version of quantum mechanics \cite{Paredes2015}; in this paper we go further and model luminous matter as an N-body system with gravitational interaction among the particles.

In this paper we perform a detailed analysis of the evolution during a head-on encounter of two structures made of BEC dark matter and luminous matter. The BEC dark matter is ruled by the typical GPP system of equations whereas the luminous matter is modeled with a system of N-particles interacting gravitationally and subject to the gravitational potential sourced by the BEC dark matter. In this paper we do not consider the back reaction of the luminous matter onto the gravitational potential, and thus we restrict to cases where the luminous matter contributes only with 5\% of the total mass of the structure. We explore the regime of solitonic behavior and actual merge, and show the implications of the encounter on the luminous matter behavior in each case.

The paper is organized as follows. In section \ref{sec:nm} we describe the GPP set of equations, the model we use for the luminous matter and show tests that have to be fulfilled by a code dealing with the GPP system plus N-particles. In section \ref{sec:results} we show the various scenarios of head-on encounters and describe some astrophysical scenarios. Finally, in \ref{sec:conclusions} we present some conclusions.

\section{Systems of equations and numerical methods}
\label{sec:nm}

\subsection{The GPP system}
\label{subsec:GPP}

The basic assumption here is that the BEC dark matter obeys the GPP system of equations. This is the constrained evolution system given by the GP equation, namely the Schr\"odinger equation for the wave function describing the gas in the mean field approximation, with a potential trap due to the gravitational field generated by the density of probability of the condensate. Such potential is in turn sourced by the density of the wave function itself through Poisson equation. Explicitly the equations are

\begin{eqnarray}
i\hbar \frac{\partial \tilde{\Psi}}{\partial \tilde{t}} &=& -\frac{\hbar^2}{2m}\tilde{\nabla}^2 \tilde{\Psi} + \tilde{V}\tilde{\Psi} +\frac{2\pi \hbar^2 \tilde{a}}{m^2} |\tilde{\Psi}|^2\tilde{\Psi}, \nonumber\\
\tilde{\nabla}^2 \tilde{V} &=& 4 \pi G m |\tilde{\Psi}|^2, \label{eq:SPcompleta}
\end{eqnarray}

\noindent where in general $\tilde{\Psi} = \tilde{\Psi}(\tilde{t},\tilde{{\bf x}})$, $m$ is the mass of the boson, $\tilde{V}$ is the gravitational potential acting as the condensate trap, $\tilde{a}$ is the scattering length of the bosons. This is a coupled system consisting of an evolution equation for $\Psi$ with a potential that is solution of Poisson equation sourced by $|\Psi|^2$.

We solve system (\ref{eq:SPcompleta}) numerically in cartesian coordinates. The first step before integrating (\ref{eq:SPcompleta}) requires the remotion of constants using the following change of variables $\hat{\Psi} = \frac{\sqrt{4\pi G}\hbar}{mc^2}\tilde{\Psi}$,
$\hat{t} = \frac{mc^2}{\hbar}\tilde{t}$, $\hat{x}_i = \frac{mc}{\hbar}\tilde{x}_i$ with $x_i=(x,y,z)$,
$\hat{V} = \frac{\tilde{V}}{mc^2}$, $\hat{a} \rightarrow \frac{c^2}{2mG}\tilde{a}$, so that the numerical coefficients $\hbar,~\hbar^2/m,~2\pi \hbar^2/m^2,~4\pi Gm$ do not appear in (\ref{eq:SPcompleta}). When the boson mass $m$ is fixed, the hatted quantities get fixed and the variables in physical units can be recovered.

On the other hand, system (\ref{eq:SPcompleta}) is invariant under the transformation $ \Psi = \hat{\Psi}/\lambda^2$, $t = \lambda^2 \hat{t}$, $x_i = \lambda\hat{x}_i$, $V = \hat{V} / \lambda^2$, $a = \lambda^2 \hat{a}$, for an arbitrary $\lambda$ \cite{GuzmanUrena2004}. This invariance allows one to reduce the original system (\ref{eq:SPcompleta}) to the following set of equations in code units
 
\begin{eqnarray}
i \frac{\partial \Psi}{\partial t} &=& -\frac{1}{2}\nabla^2 \Psi + V\Psi +a|\Psi|^2\Psi \label{eq:SchroNoUnits}\\
\nabla^2 V &=& |\Psi|^2. \label{eq:PNoUnits}
\end{eqnarray}

\noindent We consider the GPP system (\ref{eq:SchroNoUnits}-\ref{eq:PNoUnits}) to be an initial value problem and restrict in this paper to the case $a=0$ considering the recent structure formation simulations restrict to this case \cite{WaveDM}. We use Schr\"odinger equation as the equation driving the evolution, whereas Poisson equation is a constraint that has to be satisfied during the evolution. The solution method is described in \cite{GonzalezGuzman2011,PanchoBEC} and basically consists in approximating both equations using finite differences. We solve (\ref{eq:SchroNoUnits}) using the method of lines with a third order Runge-Kutta integrator, and solve (\ref{eq:PNoUnits}) using a SOR algorithm. In this paper we restrict to the case of head-on collisions and therefore Poisson equation can be solved assuming the system is axially symmetric as done for rotating BEC halos on a 3D grid \cite{PanchoBEC}.

{\it BEC halo model. } We model each halo as a ground state equilibrium configuration of the GPP system in spherical symmetry. This configuration is known to be stable under radial and non-radial perturbations \cite{BernalGuzman2006a} and also behaves as a late-time attractor of configurations starting with density profiles out of equilibrium, both, with spherical symmetry \cite{GuzmanUrena2006} and without spherical symmetry \cite{BernalGuzman2006a}. The description of how to construct these ground state equilibrium configurations can be found in \cite{GuzmanUrena2004,GuzmanUrena2006,RuffiniBonazzolla,SeidelSuen1990,PanchoBEC}.

\subsection{The luminous matter}

We model the luminous matter with a configuration of $N$ particles that interact gravitationally with each other and with the BEC dark matter. At this stage we do not consider all the ingredients used to model a state of the art luminous component, that may involve the interaction with radiation and gas, however including the gravitational interaction is already a step beyond the most recent analysis in the context of BEC dark matter \cite{Paredes2015}. 

The analysis of the behavior of luminous matter in a collision may incorporate different morphological classes, for instance disks or elliptic profiles. In our analysis, since the halos are assumed to be spherically symmetric and want to track the evolution of luminous particles without a particular morphology, we assume the luminous matter has an initially spherically symmetric distribution. For this we consider they satisfy a Plummer density profile and have velocities ensuring that the initial configuration remains nearly stationary. The implementation of the initial data is obtained following the methods explained in \cite{PlummerModel}. Later on, we describe the numerical evolution of the luminous matter using an $N-$body code with a SPH implementation to compute the density $\rho$.
Let us begin with the description of the initial data.

\subsubsection{Initial Data}

We start with $N$ particles of mass $M_L/N$, where $M_L$ is the mass of the luminous matter of the structure. These particles are distributed in such a way that they satisfy the Plummer density profile given by the expression

\begin{equation}
\label{eq:density}
\rho(r) = \frac{M_L}{4\pi}\frac{3r_a^2}{(r^2+r_a^2)^{5/2}} ,
\end{equation}

\noindent obtained from the potential

\begin{equation}
\label{eq:potential}
\Phi(r) = \frac{-GM_L}{(r^2+r_a^2)^{1/2}} ,
\end{equation}

\noindent corresponding to a central field. To do that, we start with the cumulative mass distribution

\begin{equation}
\label{eq:mass0}
m(r) = \int_0^r 4\pi r^2 \rho(r) dr,
\end{equation}

\noindent and using Eq. (\ref{eq:density}) we integrate Eq. (\ref{eq:mass0}) 

\begin{equation}
\label{eq:mass1}
m(r) = r^3(r^2+1)^{-3/2} \, ,
\end{equation}

\noindent then inverting Eq. (\ref{eq:mass1}) for $r$ we obtain

\begin{equation}
\label{eq:r}
r(m) = (m^{-2/3}-1)^{-1/2} \, .
\end{equation}

\noindent The procedure to distribute the particles is the following: for every particle we need to find the coordinates $r, \theta$ and $\phi$ of the particle's position. We choose a random number $m$ such that $0\leq m \leq M_L$ and identify it  with the cumulative mass contained within the radius $r$ for that particle. Using Eq (\ref{eq:r}) we determine the coordinate $r$ for the position of the particle. To determine the angles $\theta$ and $\phi$ we choose two random numbers. The first one between $-1$ and $1$ corresponding to the value of $\arccos(\theta)$ and the second one between $0$ and $2\pi$ corresponding to the value of $\phi$.

Now, we need to choose the magnitude and direction for the velocity of the particle with respect to the center of the distribution. First, the maximum velocity allowed for a particle at a given $r$ is the escape velocity $v_e$. This velocity can be obtained equating the potential and kinetic energies and solving for $v_e$ to obtain $v_e(r)=\sqrt{2GM_L}(r^2+r_a^2)^{-1/4}$.

Next, the probability $g(v)$ to have a velocity $v=||\vec{v}||$ at a radial position $r=||\vec{r}||$ is given by
\begin{equation}
g(v)dv \varpropto (-E(r,v))^{7/2} v^2 dv \,.
\end{equation}

\noindent In terms of the escape velocity and defining $q=v/v_e$, the distribution function for $v$ becomes:

\begin{equation}
\label{eq:g}
g(q) = (1-q^2)^{7/2} q^2 \,\,\, {\rm with} \,\,\, 0\leq q \leq 1 \,.
\end{equation}

\noindent Instead of inverting Eq.(\ref{eq:g}) to estimate $v$, we use a rejection technique. That is, choose a random number $x$ between 0 and 1 and a random number $y$ between 0 and $g_{max}=0.1$. If $y<g(x)$ we accept the pair $(x,y)$ and compute the velocity as $v=\sqrt{2GM}(x^2+r_a^2)^{-1/4}$, otherwise, repeat the process choosing another pair $(x,y)$.

In this way the particles have randomized peculiar velocities around the center and the configuration is nearly stationary.

\subsubsection{Evolution method}

We use a N-body code to compute the positions and velocities of the luminous matter during the evolution. In every step of time, Newton's equations are integrated for every particle using the acceleration produced by all the other particles and by the gravitational potential due to the BEC halos.

The density of the particles is computed using a basic SPH technique, specifically a cubique spline kernel $W$. A sum over all the particles is used to compute the density of Luminous matter

\begin{equation}
\rho_L(\vec{x}_i) = \sum_{j=1}^N m_j W(|\vec{x}_i - \vec{x}_j|,h_i) \, ,
\end{equation}

\noindent where $m_j=M_L/N$ is the mass of the $j-$th particle and $h_i$ is the smoothing length of the $i-$th particle.

\subsection{Tests}
\label{sbsec:tests}

A ground state equilibrium configuration of the GPP system is stable \cite{GuzmanUrena2006} and can be evolved basically during arbitrary long time with a 3D code as shown in \cite{PanchoBEC}. On the other hand, the ball of luminous matter has been constructed in such a way that it is approximately virialized and remains nearly unchanged in time within the BEC potential well. Nevertheless it is necessary to show that these two components can actually coexist together during long enough time. By enough time we mean the time it will take two structures to collide for the various numerical experiments below. During the evolution of a single structure, it is expected that the density profiles of both, the BEC and the luminous matter remain nearly unchanged and the tests are the following.

{\it Single configuration.} We first show that a single configuration can last during more than $t\sim 40$ in code units, which is of the order of the time window used during our simulations.

\begin{figure}
\includegraphics[width=4.25cm]{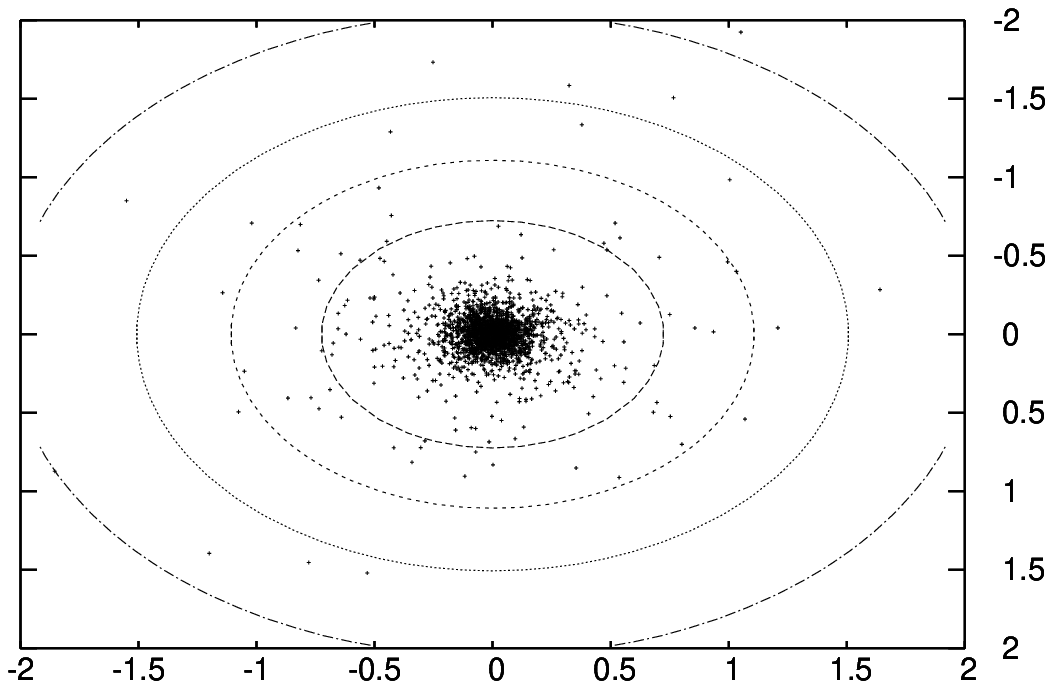}
\includegraphics[width=4.25cm]{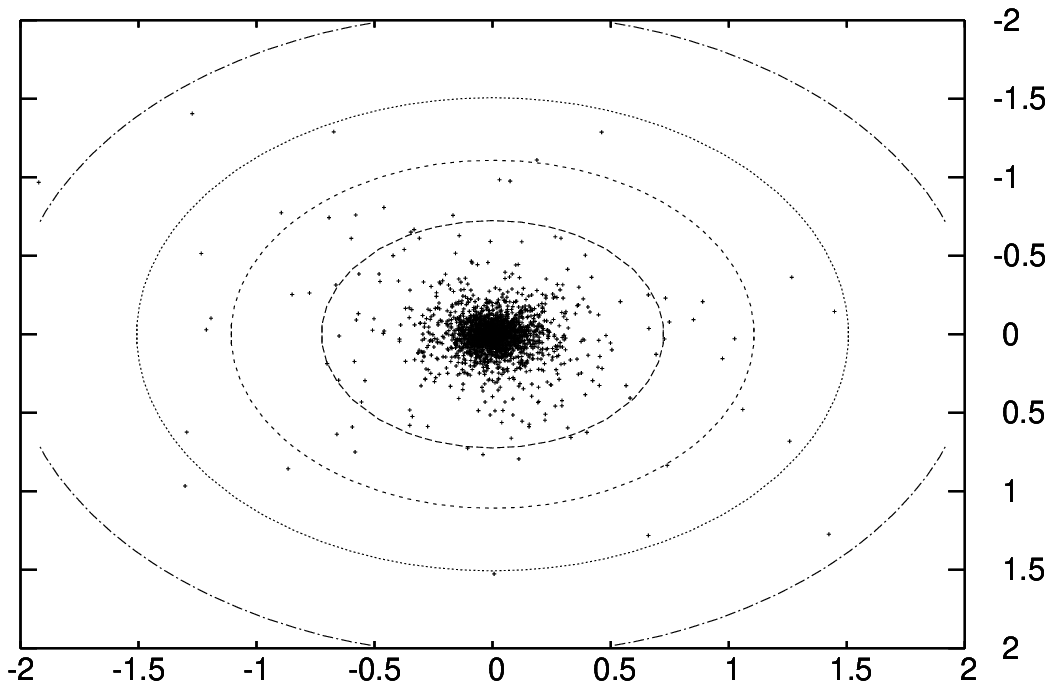}
\includegraphics[width=4.25cm]{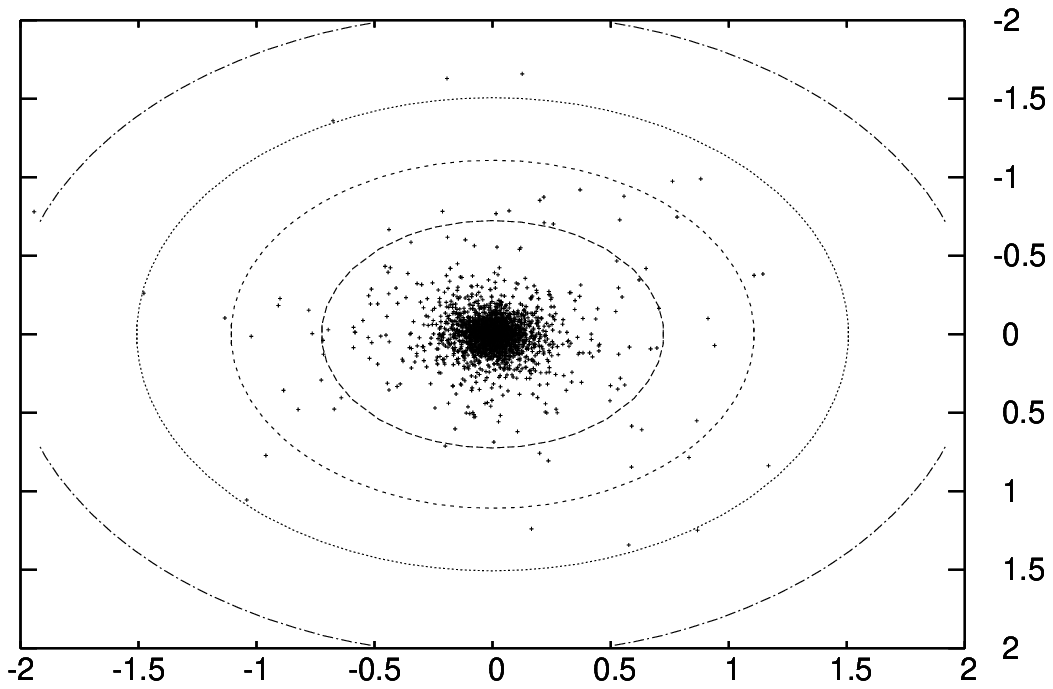}
\includegraphics[width=4.cm]{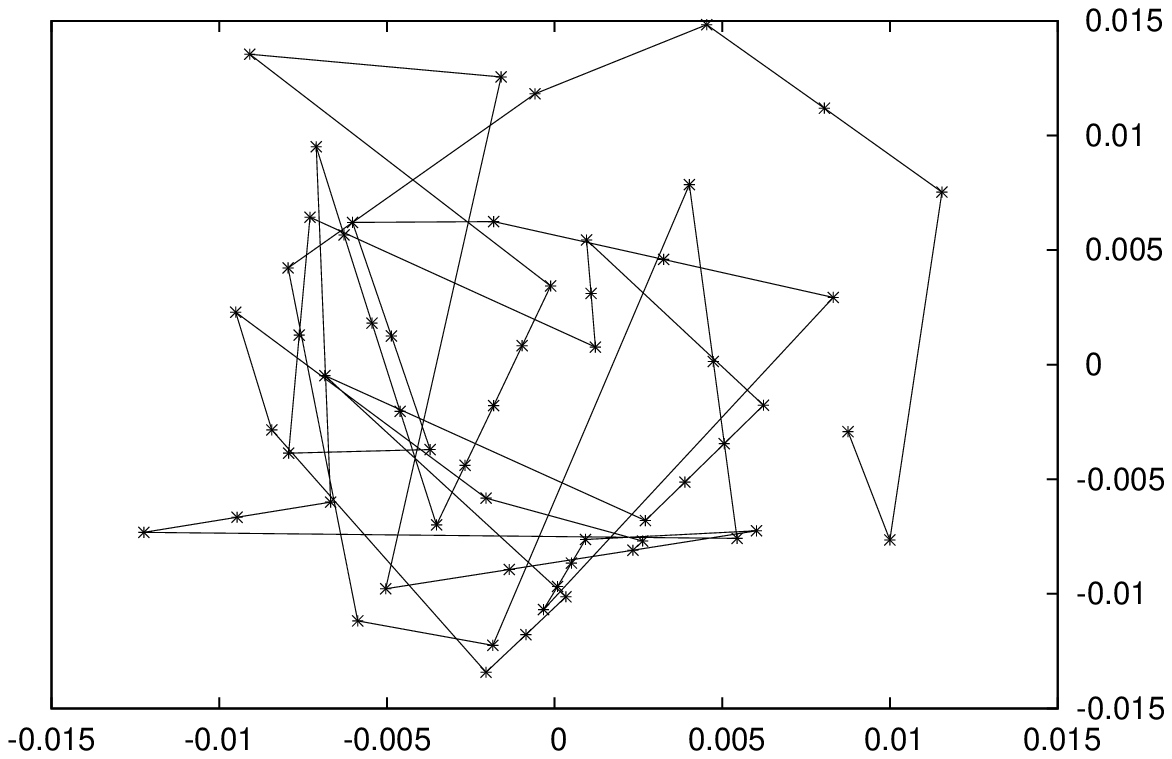}
\caption{\label{fig:test_single} We show three snapshots of the distribution of luminous particles and isocurves of $\rho_{BEC}$ at $t=0,~20,~40$ in code units. The luminous particles by construction have peculiar motion around the center of mass, however the density profile remains localized. In the bottom-right also show the position of the maximum of $\rho_{L}$ in the $xz$ plane in time, which shows the box size where the maximum of $\rho_L$ is changing.}
\end{figure}

We construct a halo+matter structure such that $M_L$ is 5\% that of the BEC halo mass, whereas the core radius of the luminous matter configuration is chosen to be $r_a = 0.25 R_{95}$, where $R_{95}$ is the radius containing 95\% of the total mass of the BEC halo. In Fig. \ref{fig:test_single} we show snapshots of the density function $\rho_{L}$ representing the luminous matter and isocurves of the density of the BEC $\rho_{BEC}=|\Psi|^2$. In the last panel we show the position of the maximum of $\rho_{L}$, which varies within a box of size $0.01r_a$. Notice that the velocities of the luminous particles is set using random numbers, thus we have checked this test for many runs, i.e. for many different initial conditions of the luminous matter.

{\it Boosted configuration.} A second test is the coexistence of the two types of matter when they are boosted. Considering $\Psi$ to be the wave function of an equilibrium ground state configuration centered at the coordinate origin, we boost it along the $z$ axis with a velocity $v_z$ by redefining $\Psi = e^{iv_z}\Psi$, provided the associated momentum $p_z$ is normalized with the mass of the BEC configuration.

The luminous matter is also boosted with the same velocity, that is, we add $v_z$ to the peculiar velocity found during the construction of the Plummer ball.

\begin{figure}
\includegraphics[width=4.25cm]{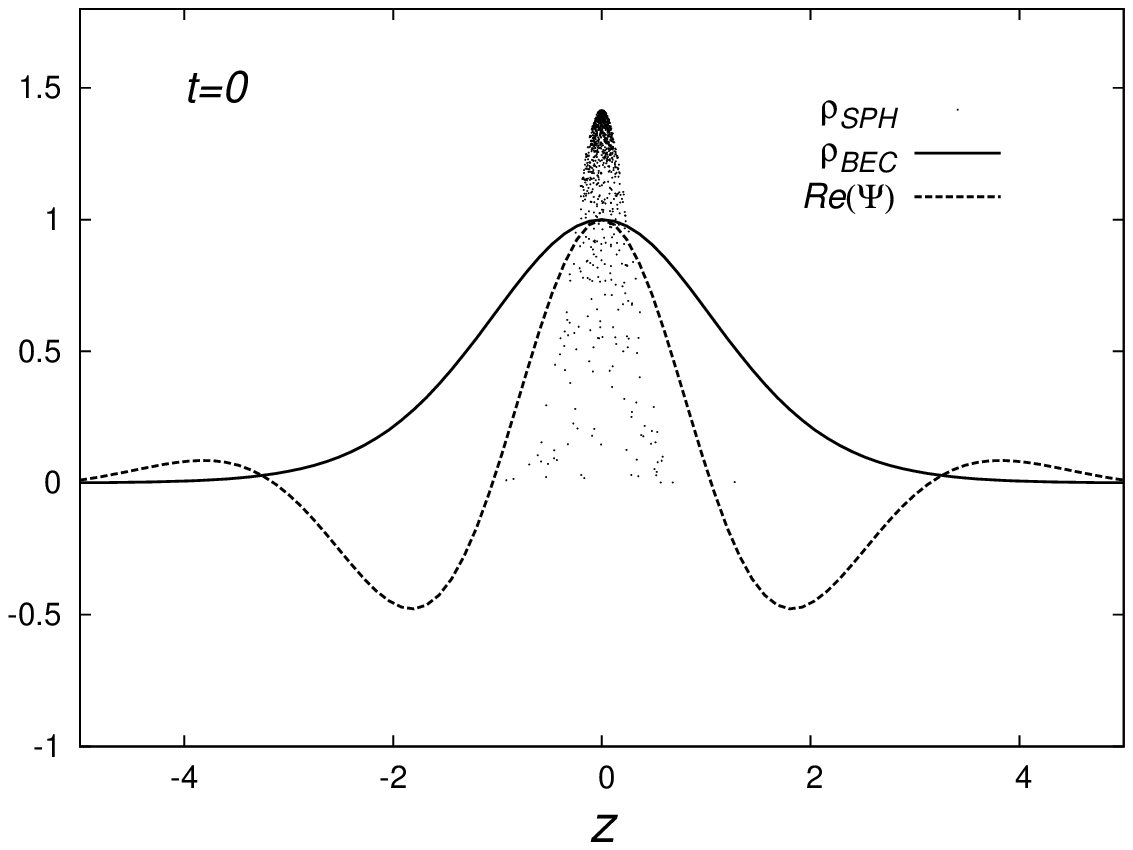}
\includegraphics[width=4.25cm]{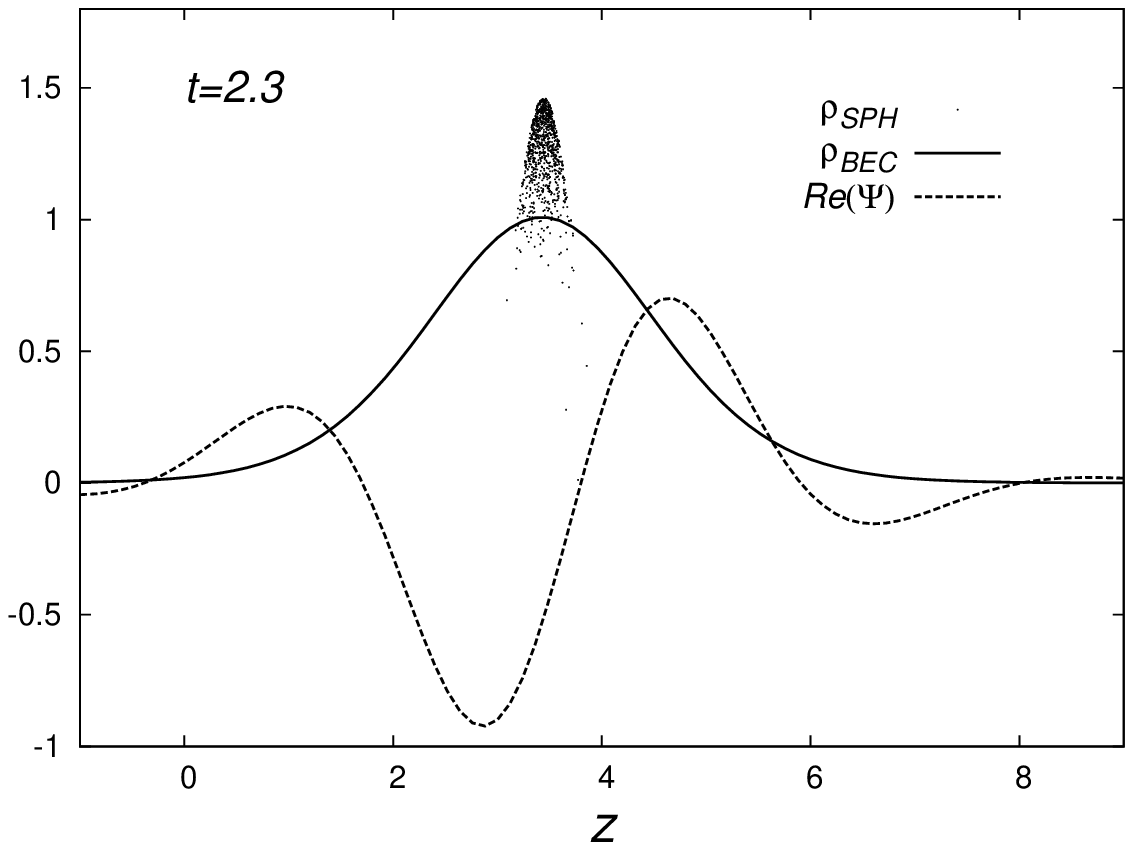}
\includegraphics[width=4.25cm]{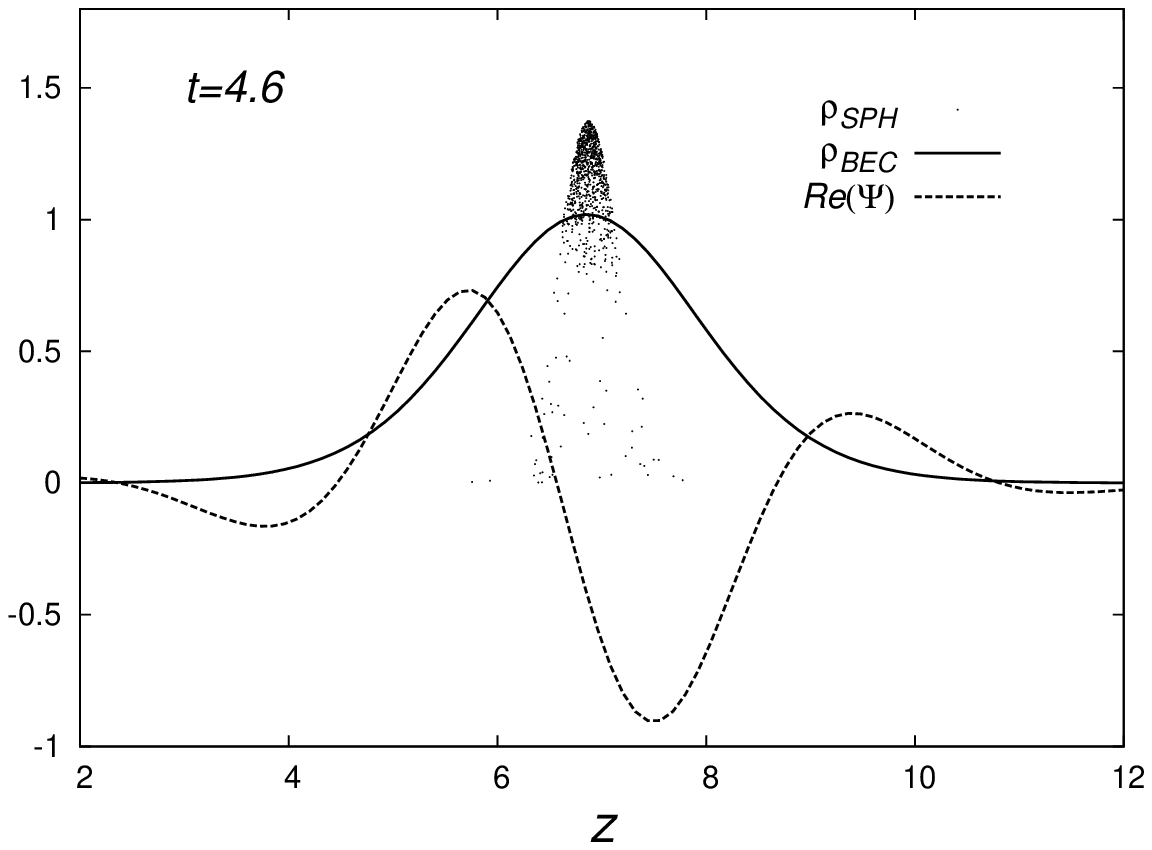}
\includegraphics[width=4.25cm]{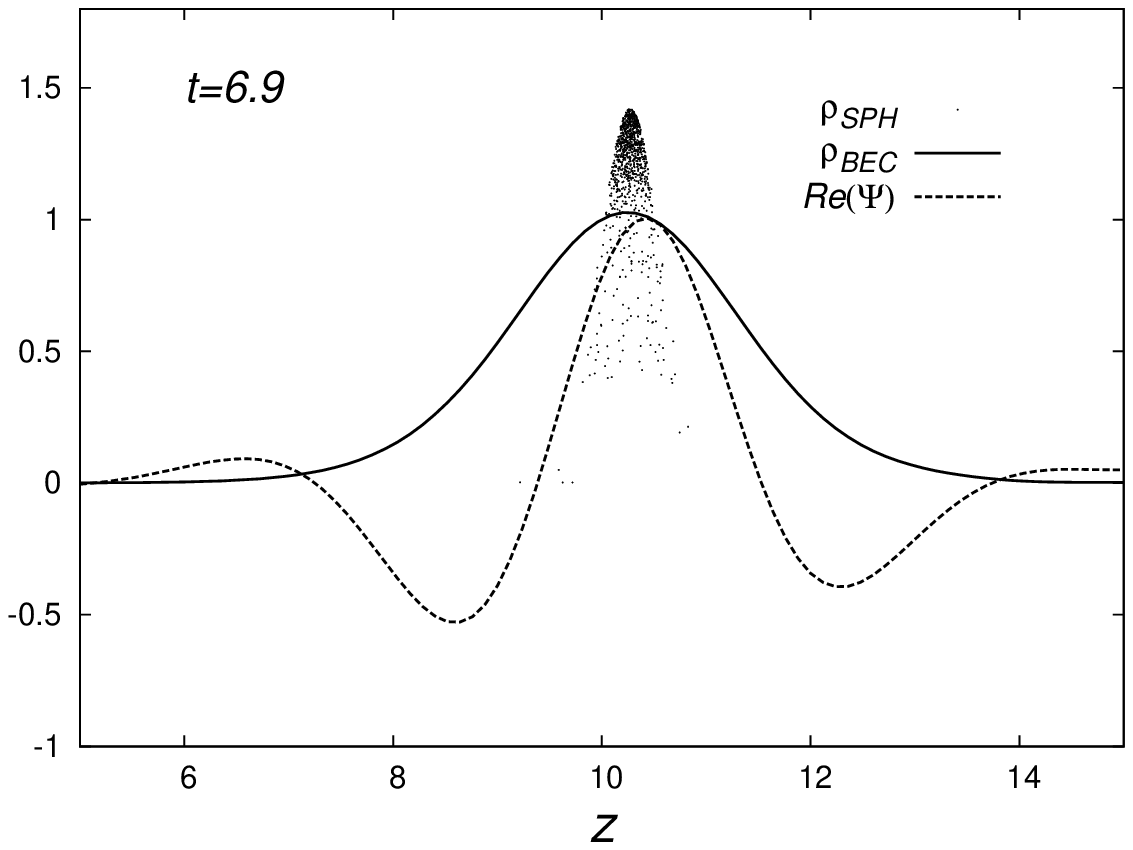}
\includegraphics[width=4.25cm]{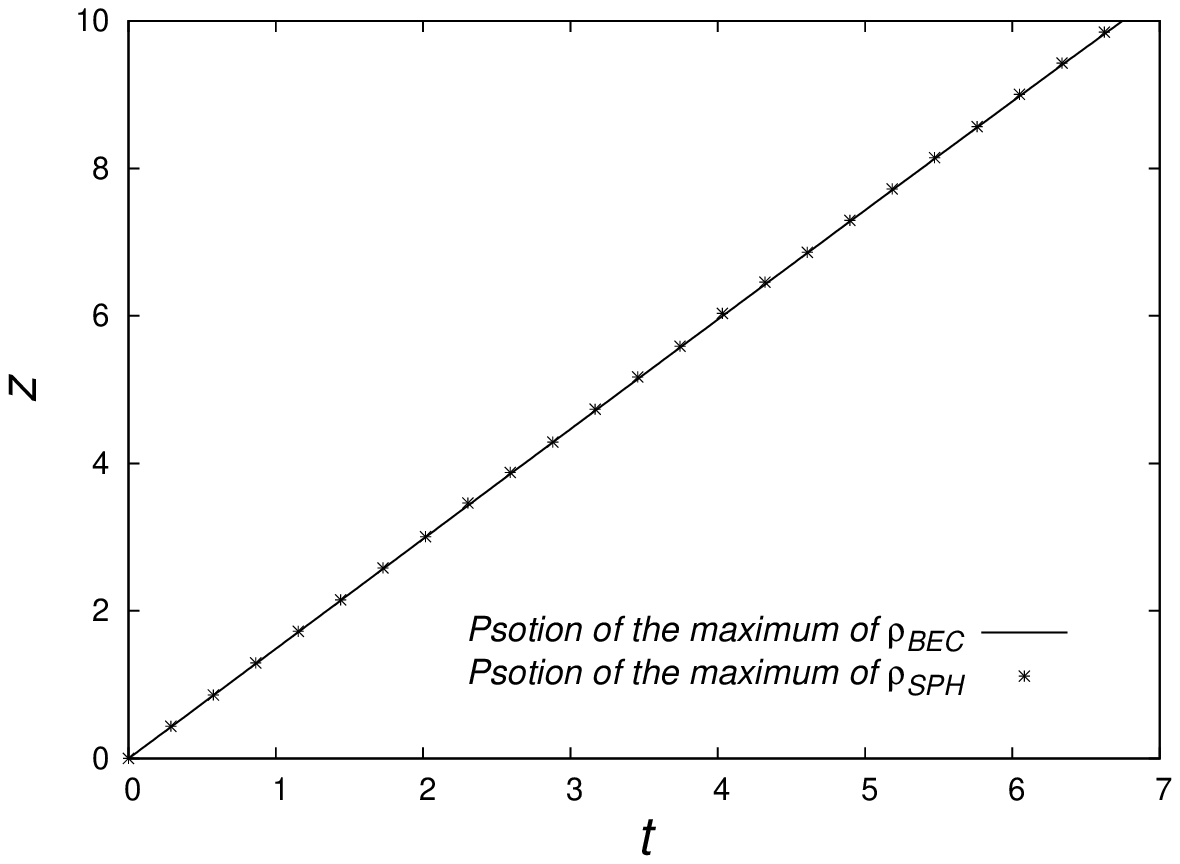}
\includegraphics[width=4.25cm]{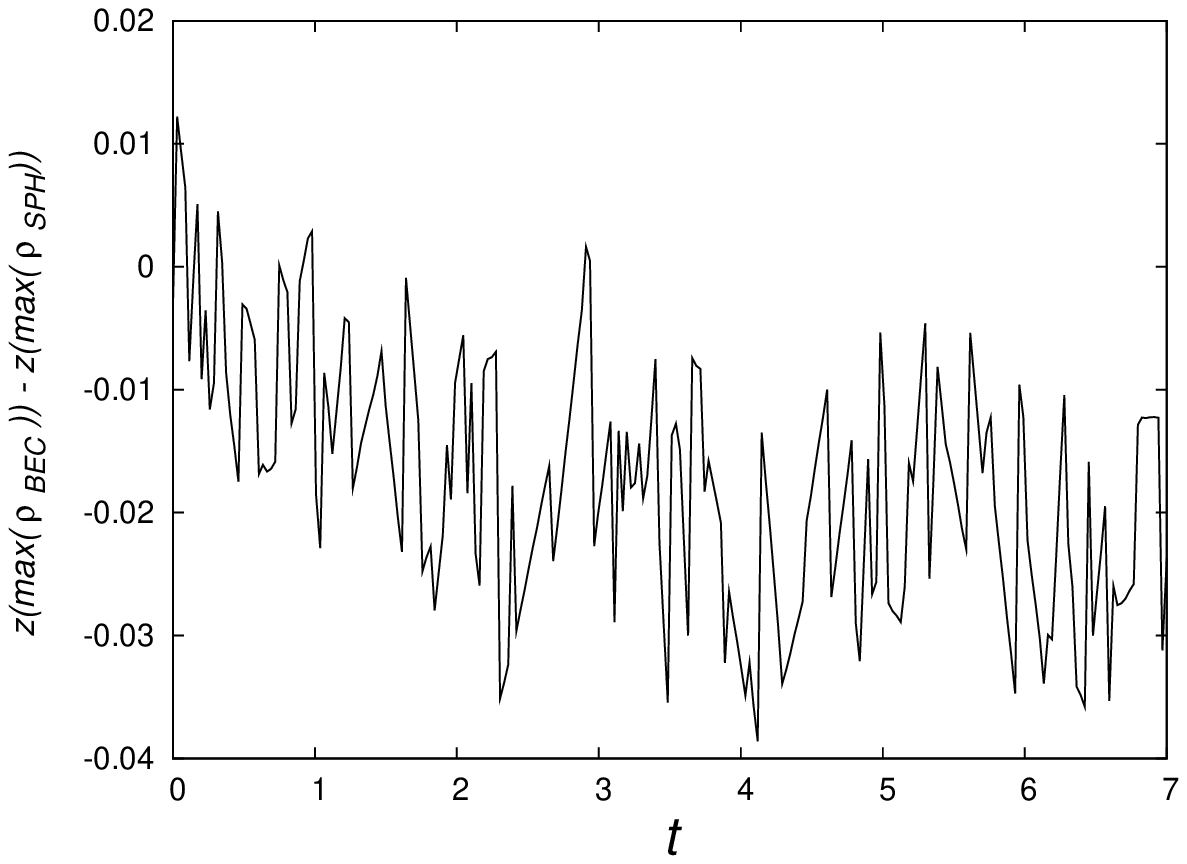}
\caption{\label{fig:test_boosted} We show four snapshots of $\rho_{BEC}$, $\rho_{L}$ and $Re(\Psi)$ at various times for a boosted configuration. The density profiles of each component remain nearly time independent as expected. The initial momentum added to the configuration is $v_z=1.5$, which is the highest we use in the paper. In the bottom panel we show the location of the maximum of both components and the difference between them along the $z$ axis, which is the direction of the boost.}
\end{figure}

The density profile of both components has to remain nearly time independent during a sufficiently long  time. In Fig. \ref{fig:test_boosted} we show snapshots of the structure with both components, BEC and luminous matter, traveling with a boost of $v_z=1.5$.  The density profile of the two components remains nearly time independent and the maxima of the two densities follow nearly the same trajectory, as expected because both, the luminous and dark matter have the same boost velocity. At the same time we show $Re(\Psi)$, which is evolving in time. In the bottom of  Fig. \ref{fig:test_boosted} we show the location of the maximum of $\rho_{BEC}$ and $\rho_L$, which should have the same trajectory in theory. In practice though, there is a drift that we measure and is of the order of $2\%$ for $v_z=1.5$, the case of highest boost in this paper. We have verified that this drift is smaller or equal for a range of the number of particles between 1000 and 10000, used to model the luminous matter in our simulations. The numerical domain used is $[-30,30]^3$ in code units in this test and our production runs.

This test is extremely important because it shows that, when a structure travels alone, it evolves correctly in the absence of other structures. This means that a behavior different from the one seen in Fig. \ref{fig:test_boosted} for each of the two configurations prior to a collision, will be due to the interaction between the two structures and not to the initial boost.

\section{Head-on encounters of two equal structures}
\label{sec:results}

\subsection{Simulations of encounters}

For the collision we set two structures of BEC and luminous matter boosted along the head-on collision direction. For this we interpolate the wave function for the BEC and the $\rho_L$ of the equilibrium configuration constructed in \ref{sbsec:tests} at two different places on the evolution grid $(0,0,\pm z_0)$ along the $z-$axis. 

For the initial BEC configuration we define a global wave function describing the state as $\Psi=\Psi_1+\Psi_2$, where $\Psi_1$ and $\Psi_2$ represent the two configurations placed in the domain at $(0,0,- z_0)$ and $(0,0,+ z_0)$ respectively. We choose $z_0$ such that the two configurations are far enough one from the other so that the interference $\langle\Psi_1,\Psi_2\rangle$ is near to round off error values as done in \cite{BernalGuzman2006b}. In this way, we have initial data for two superposed ground state equilibrium configurations in our domain.

After this, we apply momentum along the head-on direction, namely the $z$ axis. For this we simply proceed as in the boosted configuration above and define a new wave function $\Psi=e^{iv_z}\Psi_1 + e^{-iv_z}\Psi_2$. Once this phase correction is applied to the wave function of the BEC we solve Poisson equation to complete the preparation of the initial state. As done for the boosted configurations above, we also add linear momentum $v_z$ to all the luminous particles along the head-on collision direction at initial time.

\begin{figure}
\includegraphics[width=4.25cm]{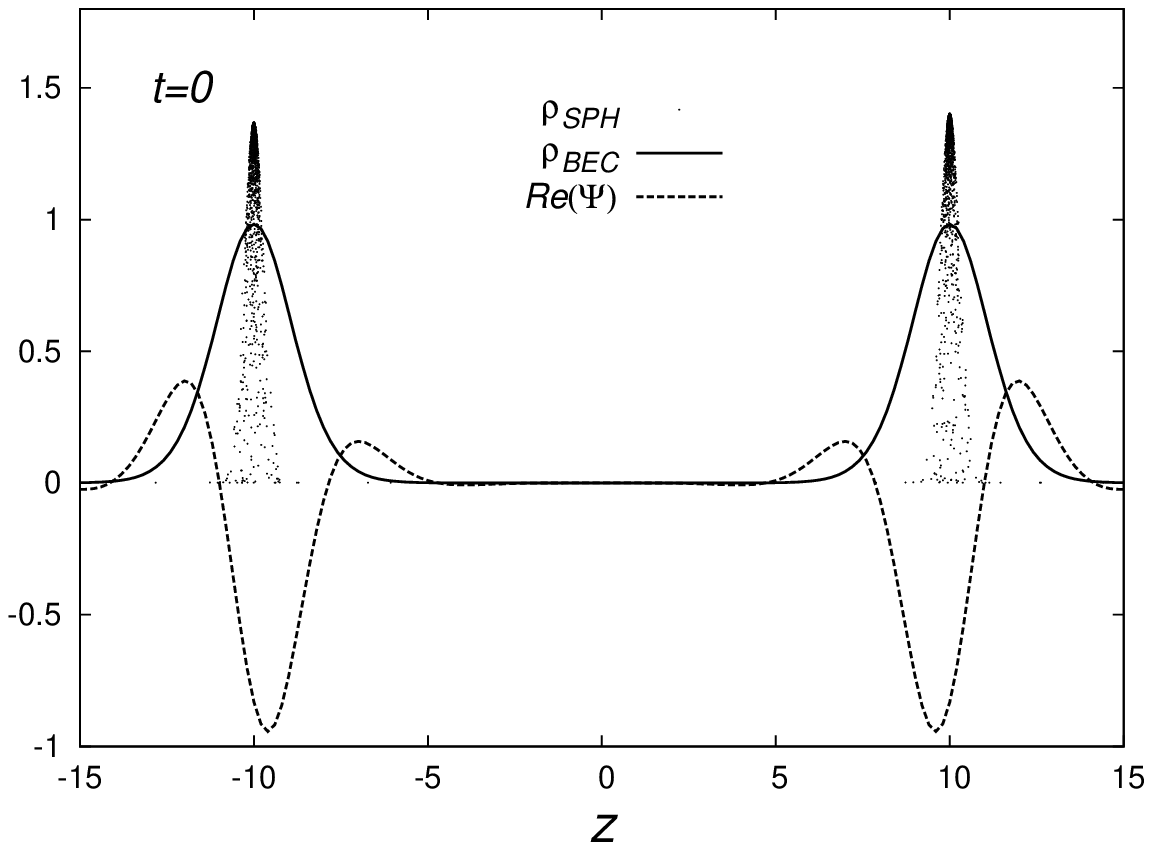}
\includegraphics[width=4.25cm]{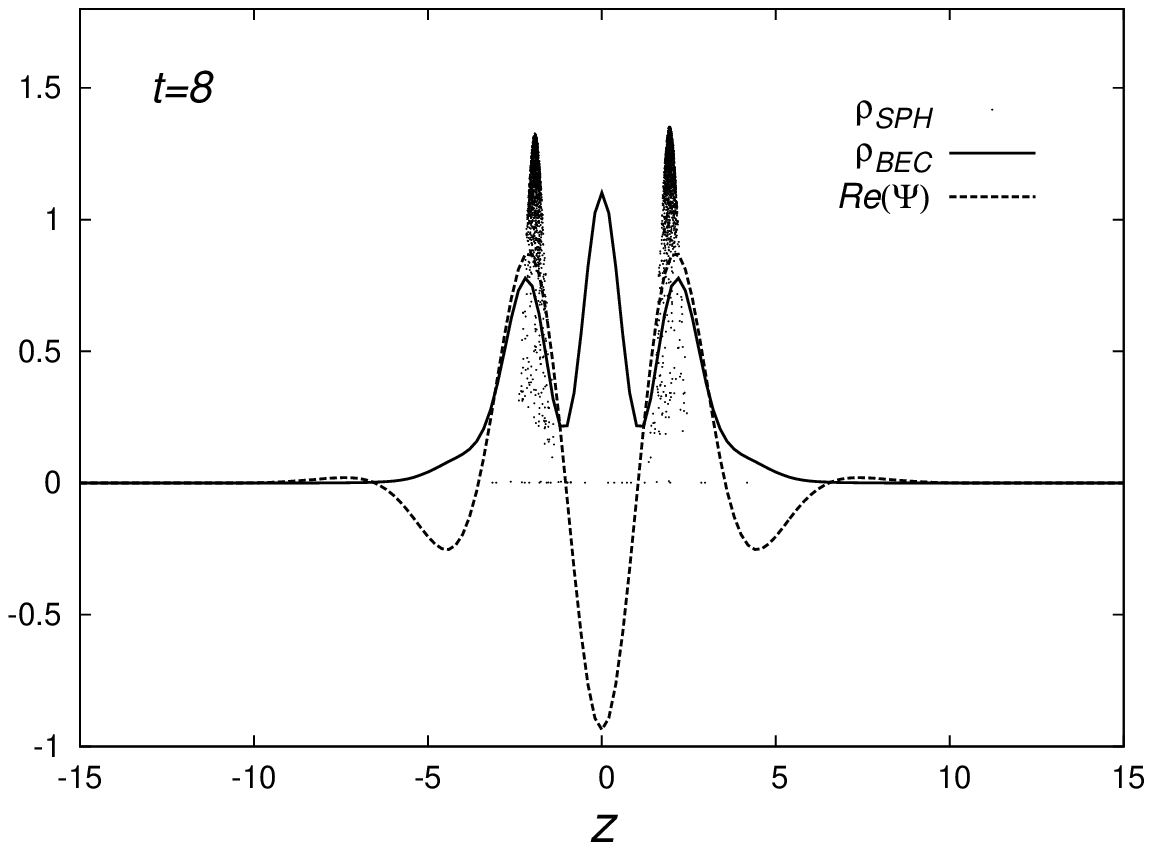}
\includegraphics[width=4.25cm]{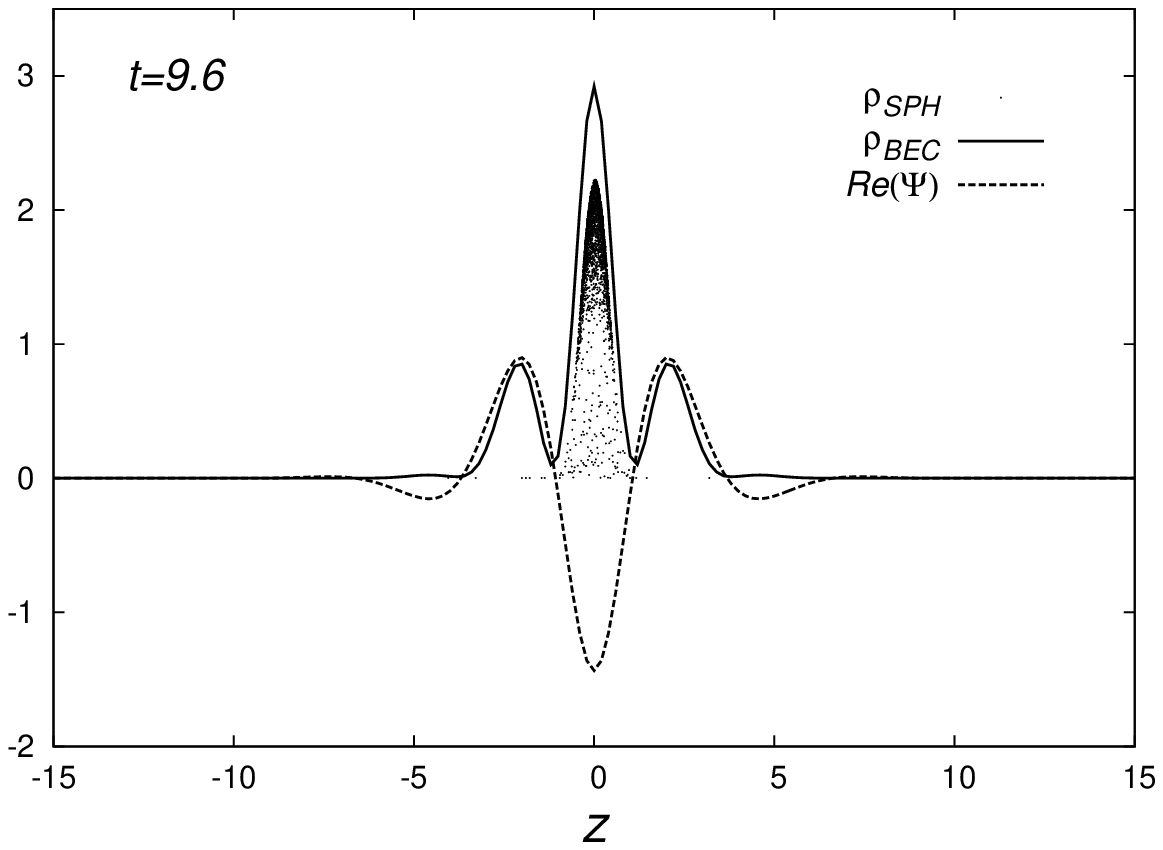}
\includegraphics[width=4.25cm]{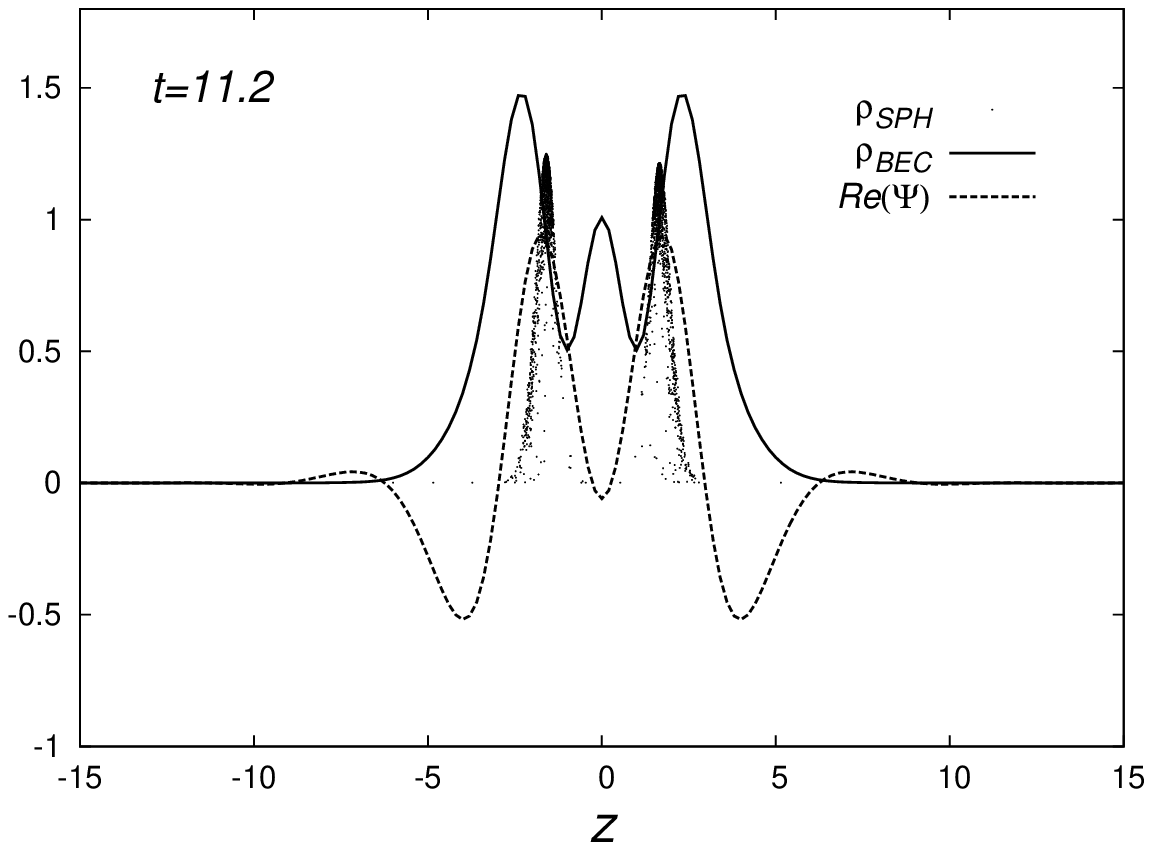}
\includegraphics[width=4.25cm]{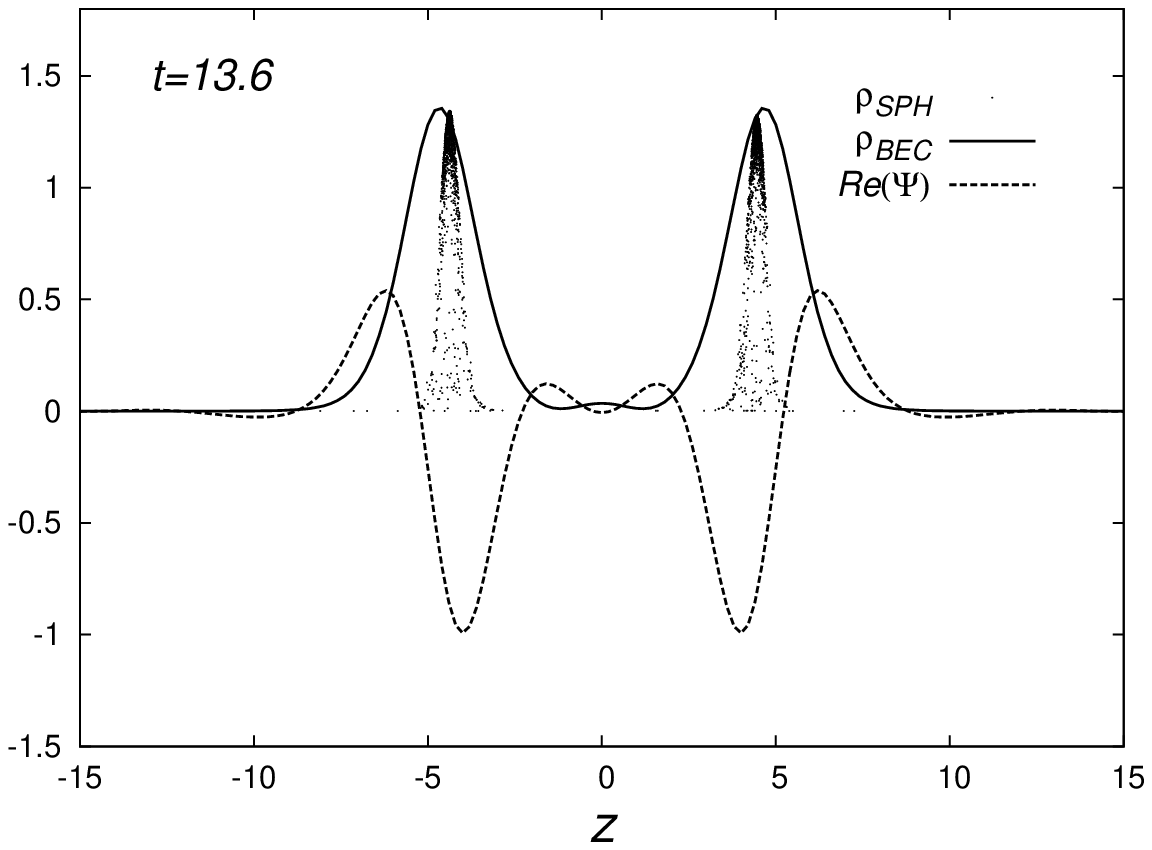}
\includegraphics[width=4.25cm]{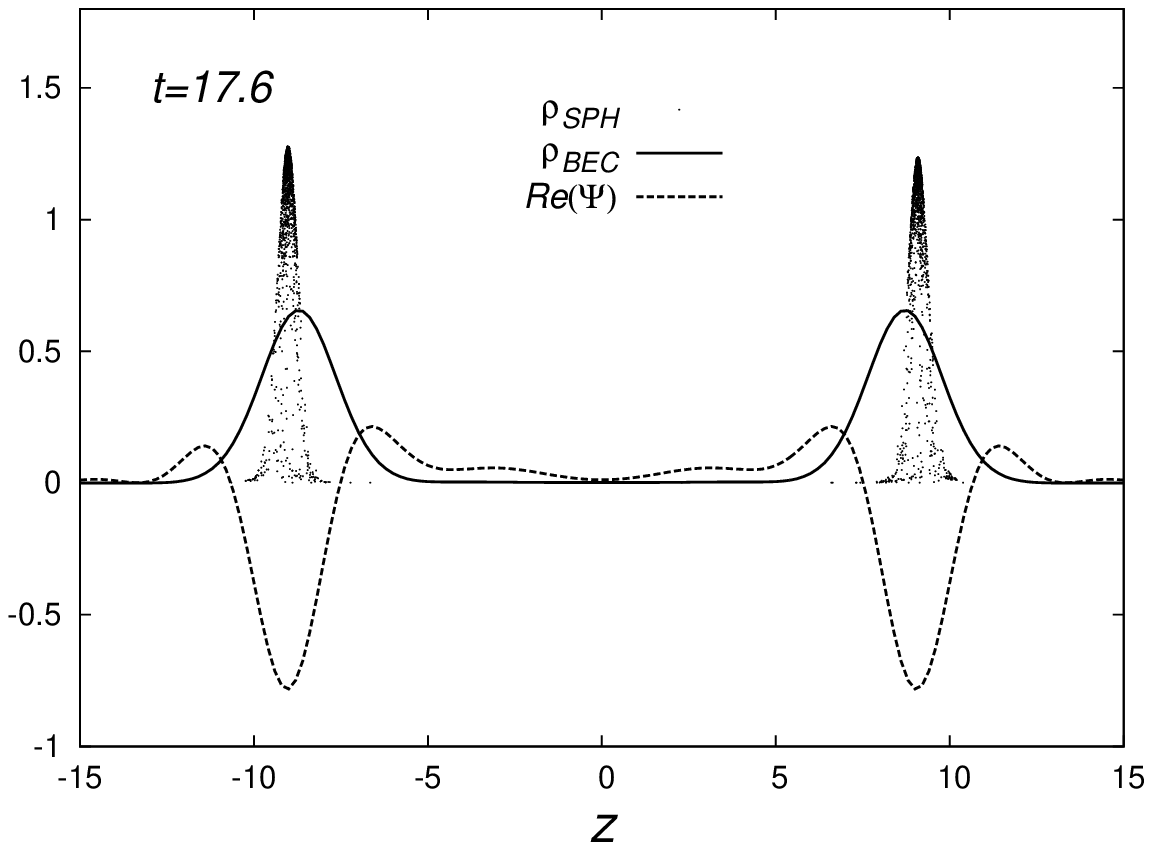}
\caption{\label{fig:collision} We show snapshots of $\rho_{BEC}$, $\rho_{L}$ and $Re(\Psi)$ at various times for a collision with $v_z=1$. In the bottom panel we show the trajectory of the maximum of each blob for dark and luminous matter. We use 5000 particles for the luminous matter.}
\end{figure}

Then we start up the evolution. As illustration we show in Fig. \ref{fig:collision} various snapshots of the density profile of luminous matter, dark matter and $Re(\Psi)$ for the case $v_z=1$. The different snapshots show how before the collision the blobs proceed as in the boosted case. When the system is near to collide the BEC forms an interference pattern by $t=8$. During the collision of luminous matter, the interference pattern persists by $t=9.6$. Later on, after the collision, the BEC shows a solitonic behavior and the luminous matter moves behind the two blobs of BEC as can be seen in the snapshots at $t=11.2$ and $t=13.6$. By $t=17.6$ it can be seen that the luminous matter takes over the peaks of the BEC component. The domain of this and all our simulations is $[-30,30]^3$.

This particular case with $v_z=1$ shows a genuine solitonic behavior of the BEC. Recalling the results in \cite{BernalGuzman2006b}, the sign of the total energy of the BEC $E=K+W+I$, kinetic, plus gravitational plus self-interaction, determines whether or not there will be solitonic behavior. In this paper $a=0$ and therefore $I=0$. For $E<0$ the two structures of BEC get glued together and form a single potential well, whereas for $E>0$ the two BEC component trespass each other and continue toward infinity showing a nearly solitonic behavior \cite{Solitons}. The threshold is found for approximately $v_z=0.755$, which means that in some cases $E>0$ ($v_z>0.755$) and $E<0$ ($v_z < 0.755$) in others \cite{BernalGuzman2006b}.

\begin{figure}
\includegraphics[width=4.25cm]{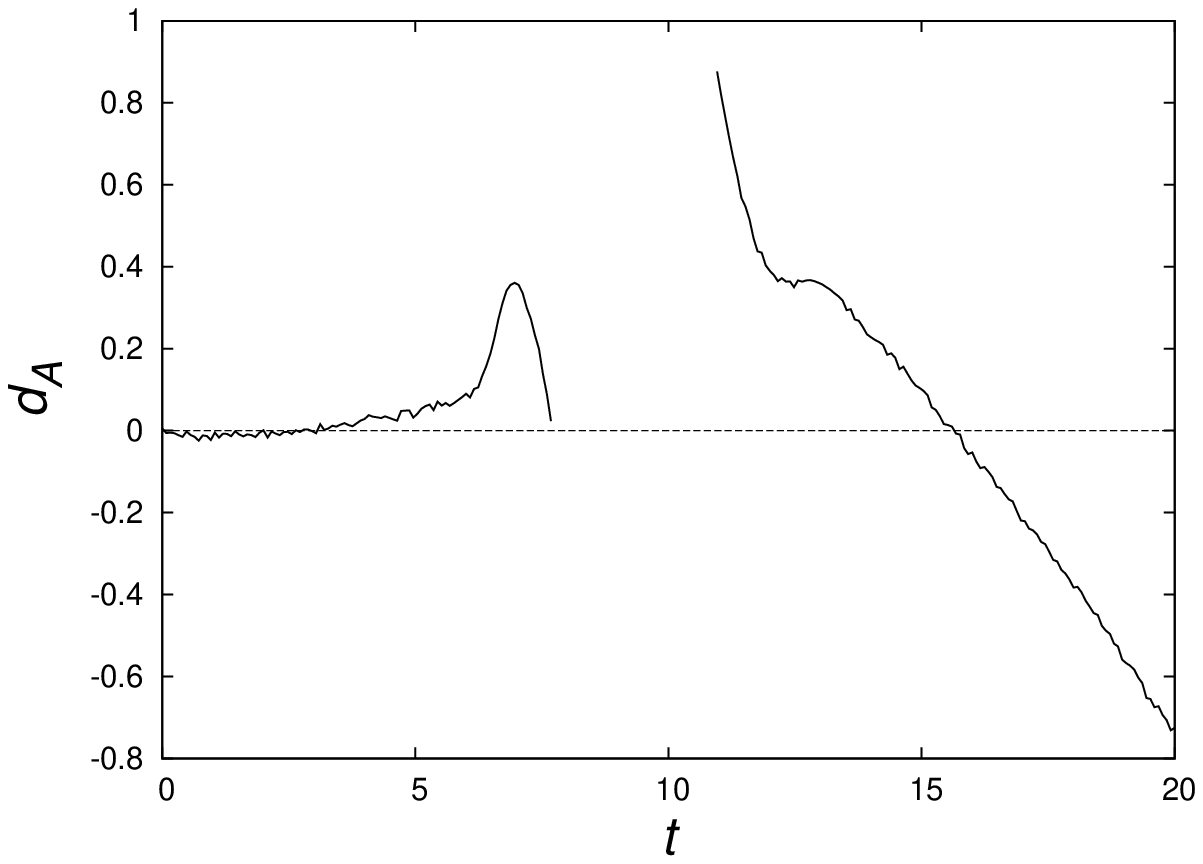}
\includegraphics[width=4.25cm]{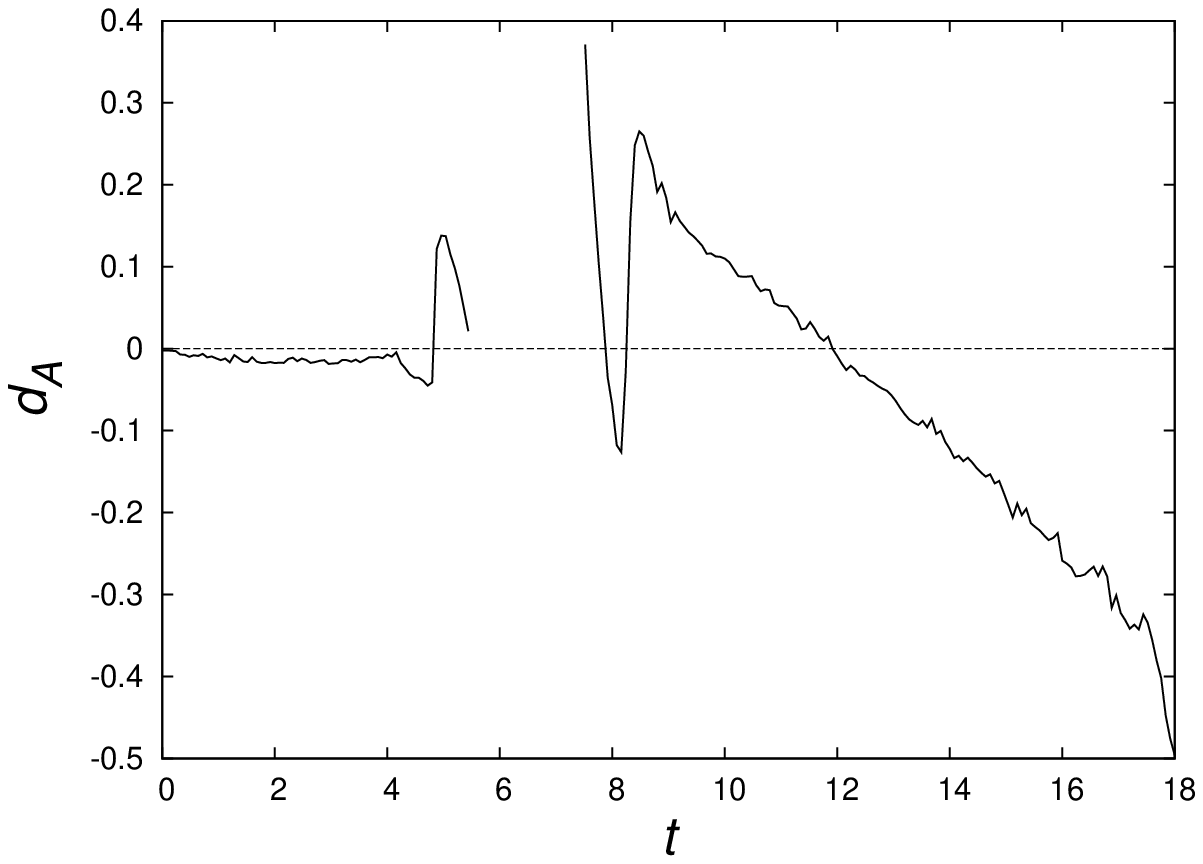}
\caption{\label{fig:d_A} Values of $d_A$ for $v_z=1$ and $v_z=1.5$ respectively. We have removed the time window where the interference pattern occurs and the maximum of $\rho_{BEC}$ is located basically at the origin.}
\end{figure}

\begin{figure}
\includegraphics[width=4.25cm]{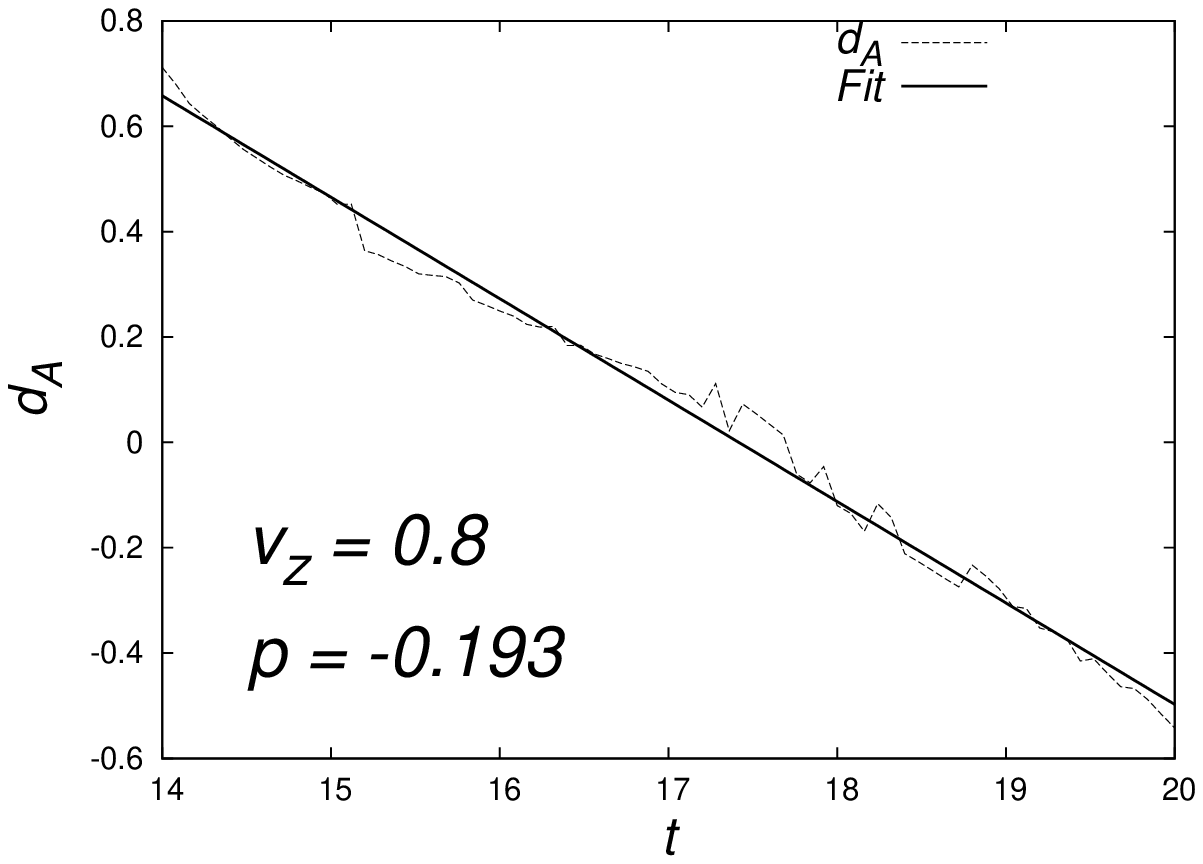}
\includegraphics[width=4.25cm]{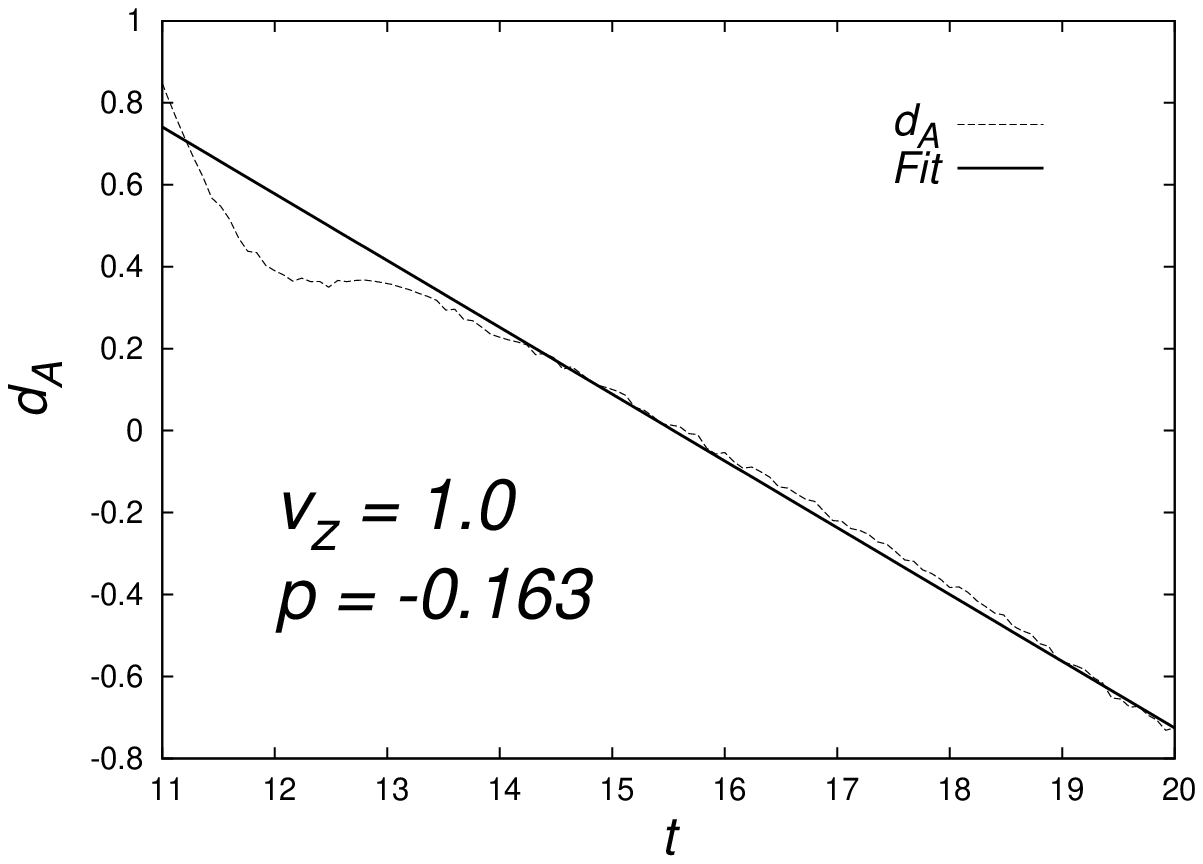}
\includegraphics[width=4.25cm]{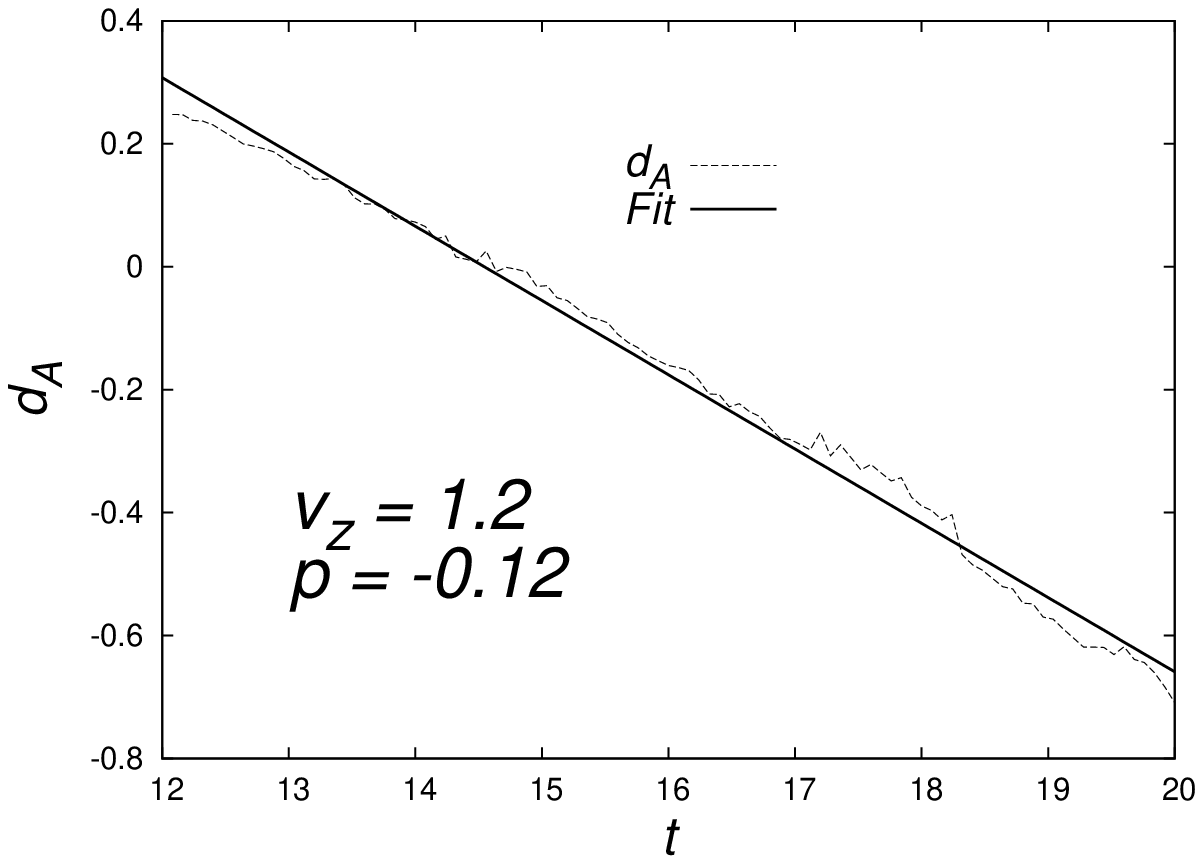}
\includegraphics[width=4.25cm]{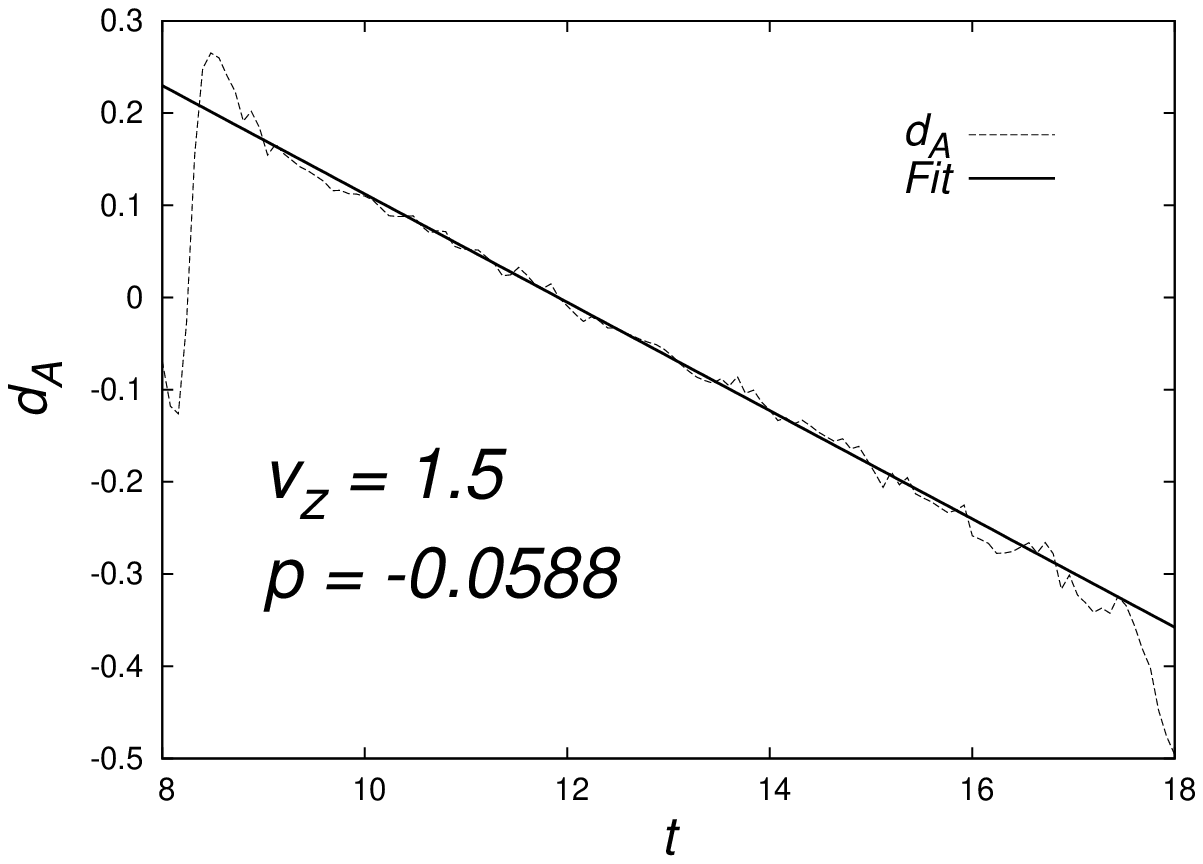}
\caption{\label{fig:fits} We show the fits of the relative motion between $\rho_{BEC}$ and $\rho_{L}$ for various values of $v_z>0.755$ after the collision.}
\end{figure}

\begin{figure}
\includegraphics[width=4.25cm]{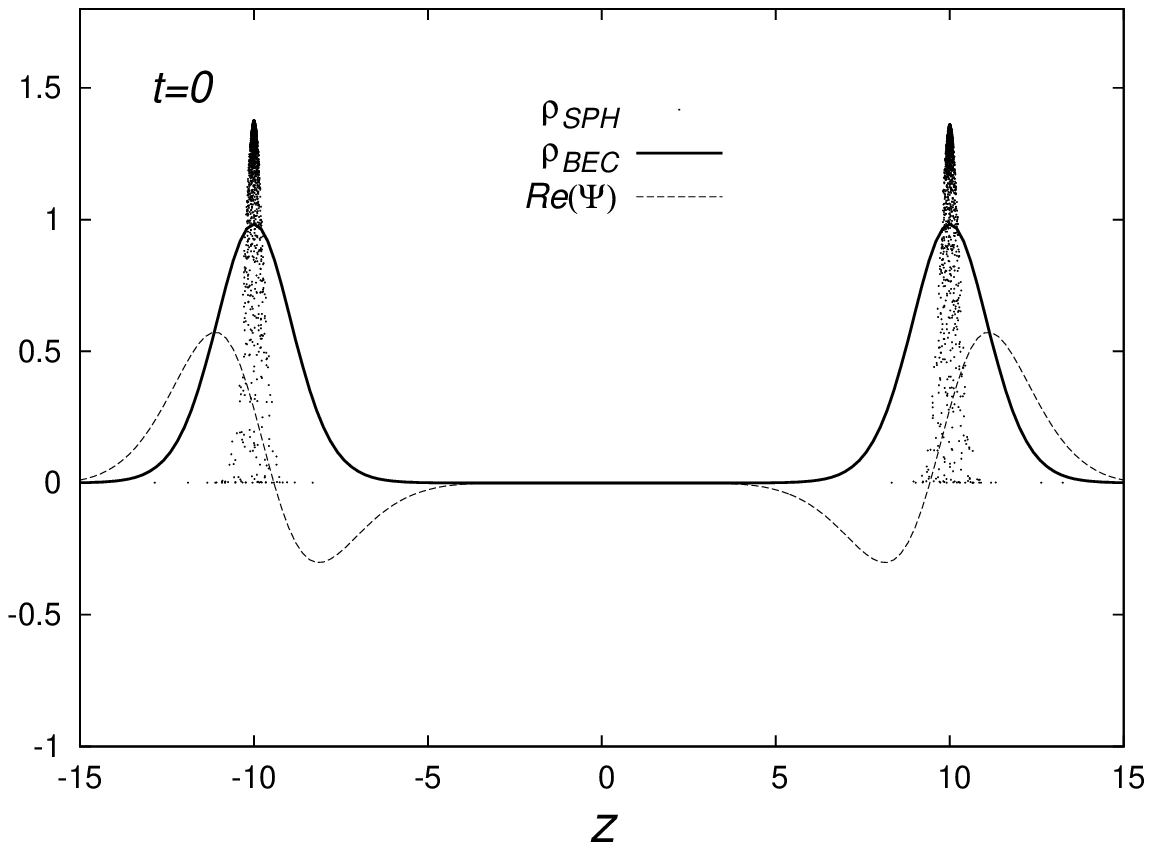}
\includegraphics[width=4.25cm]{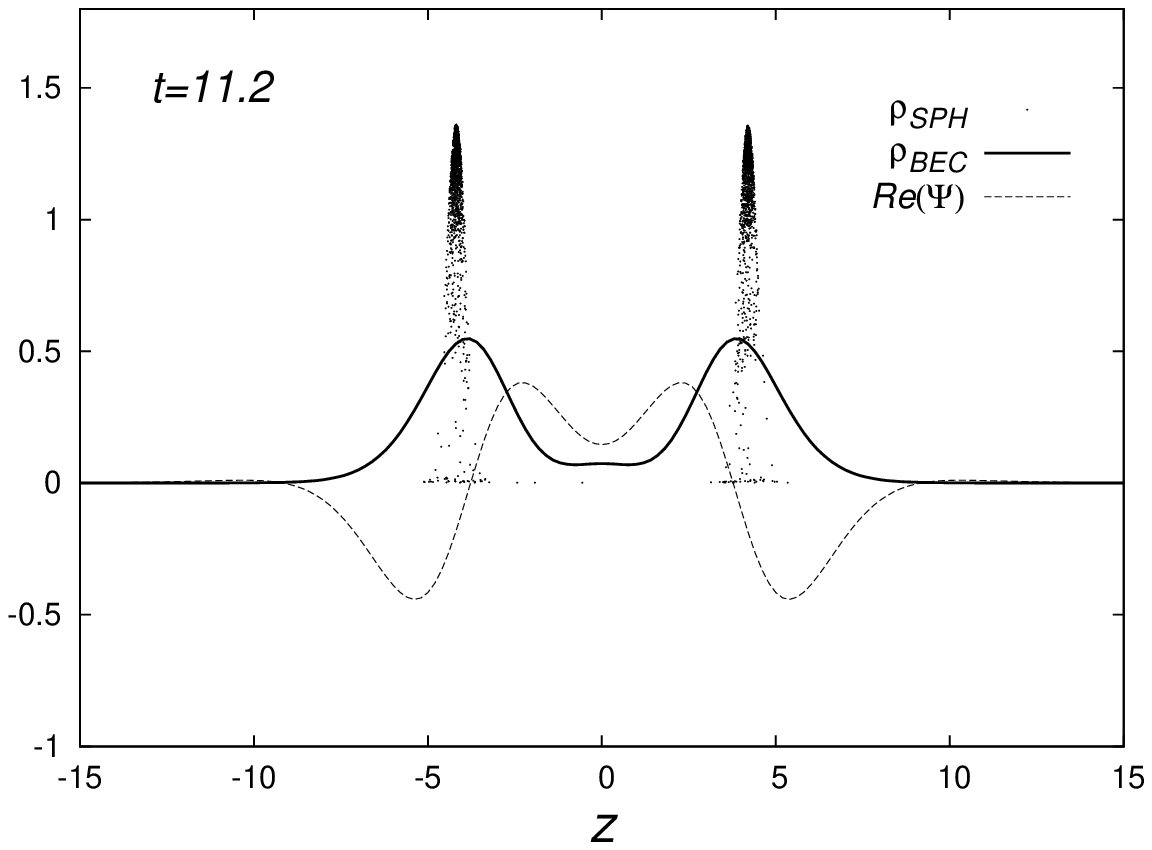}
\includegraphics[width=4.25cm]{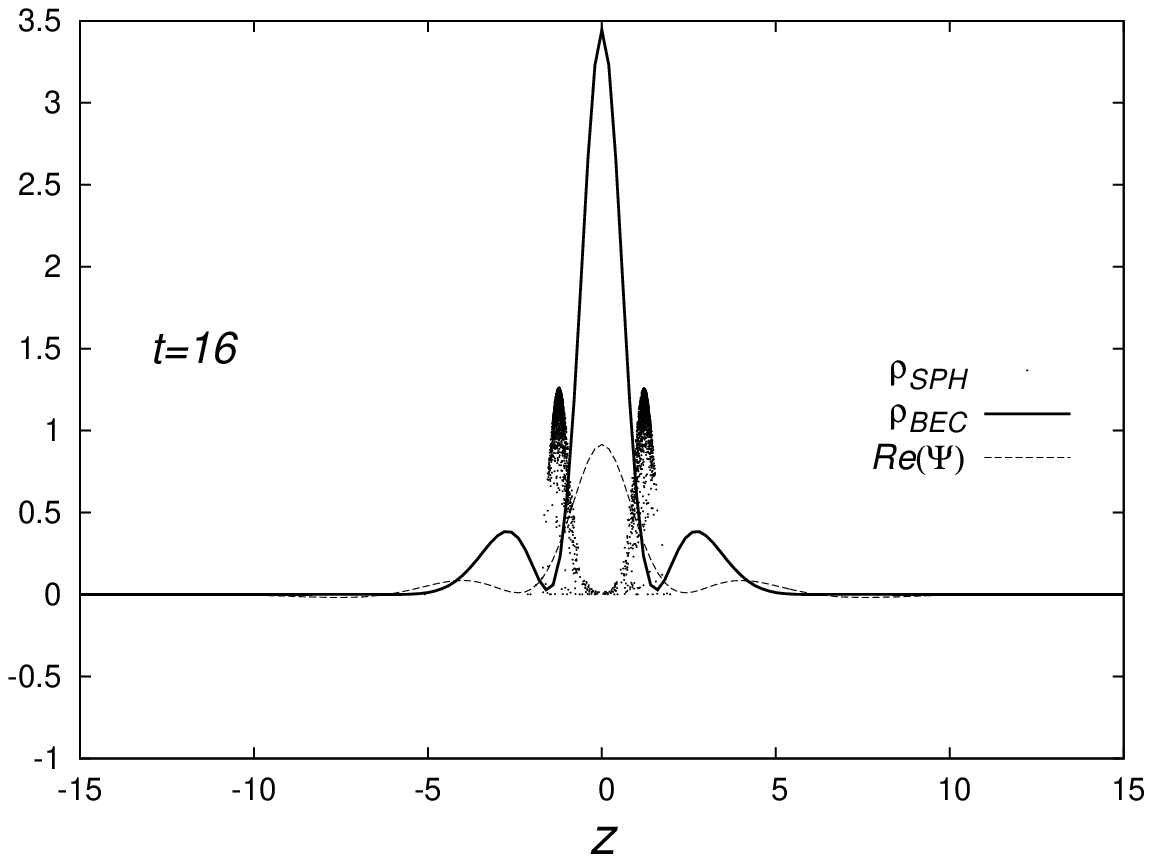}
\includegraphics[width=4.25cm]{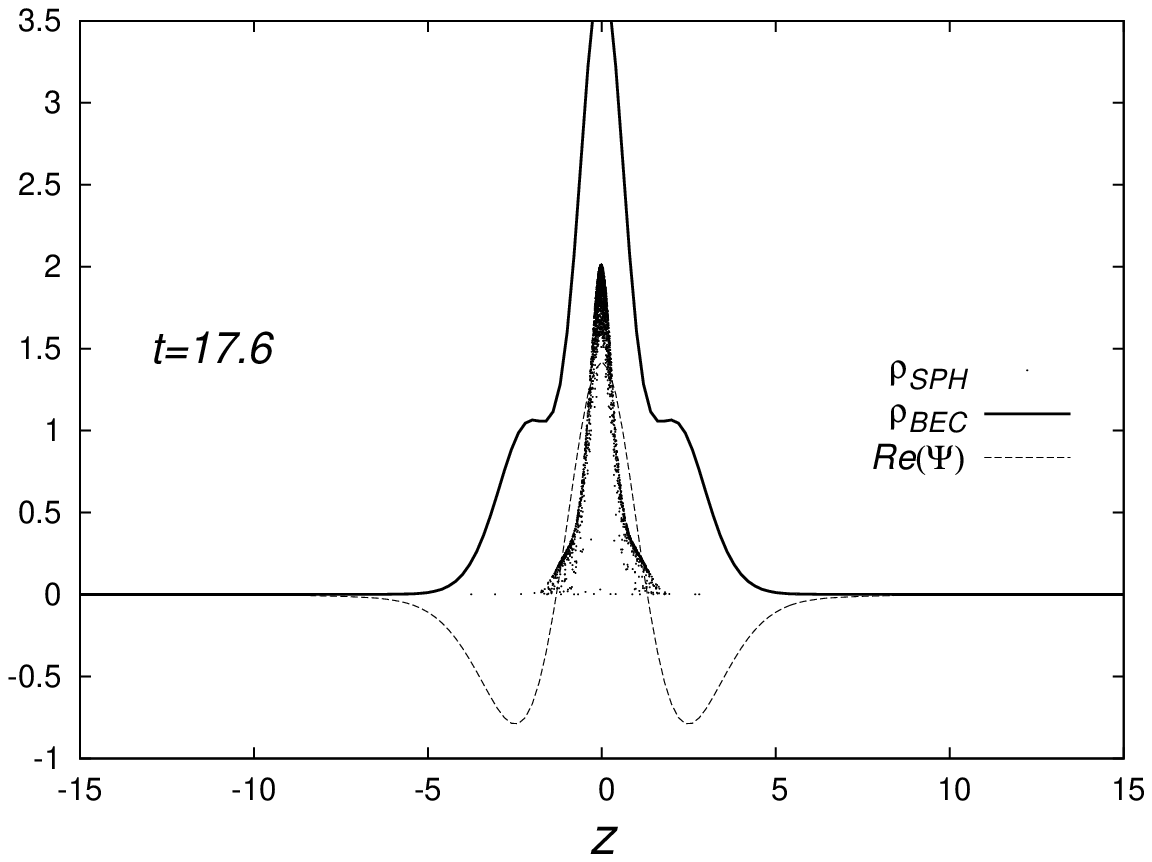}
\includegraphics[width=4.25cm]{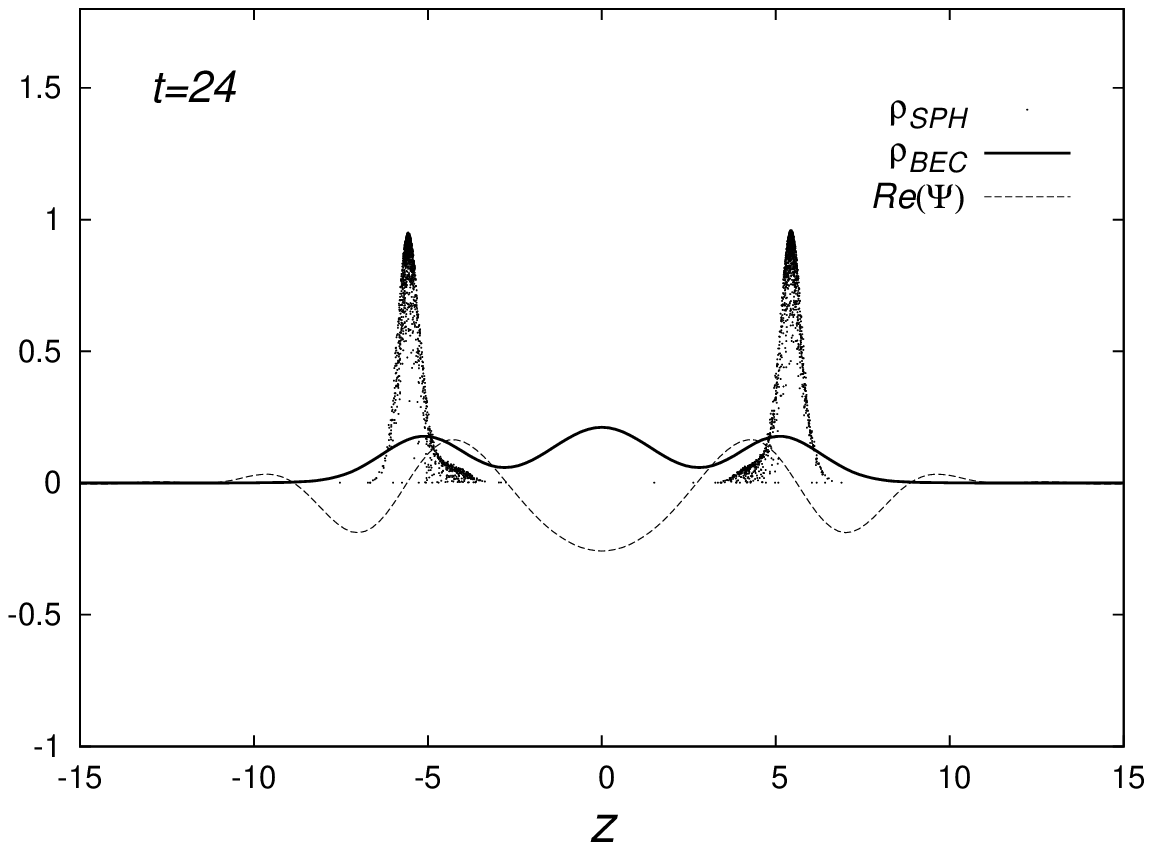}
\includegraphics[width=4.25cm]{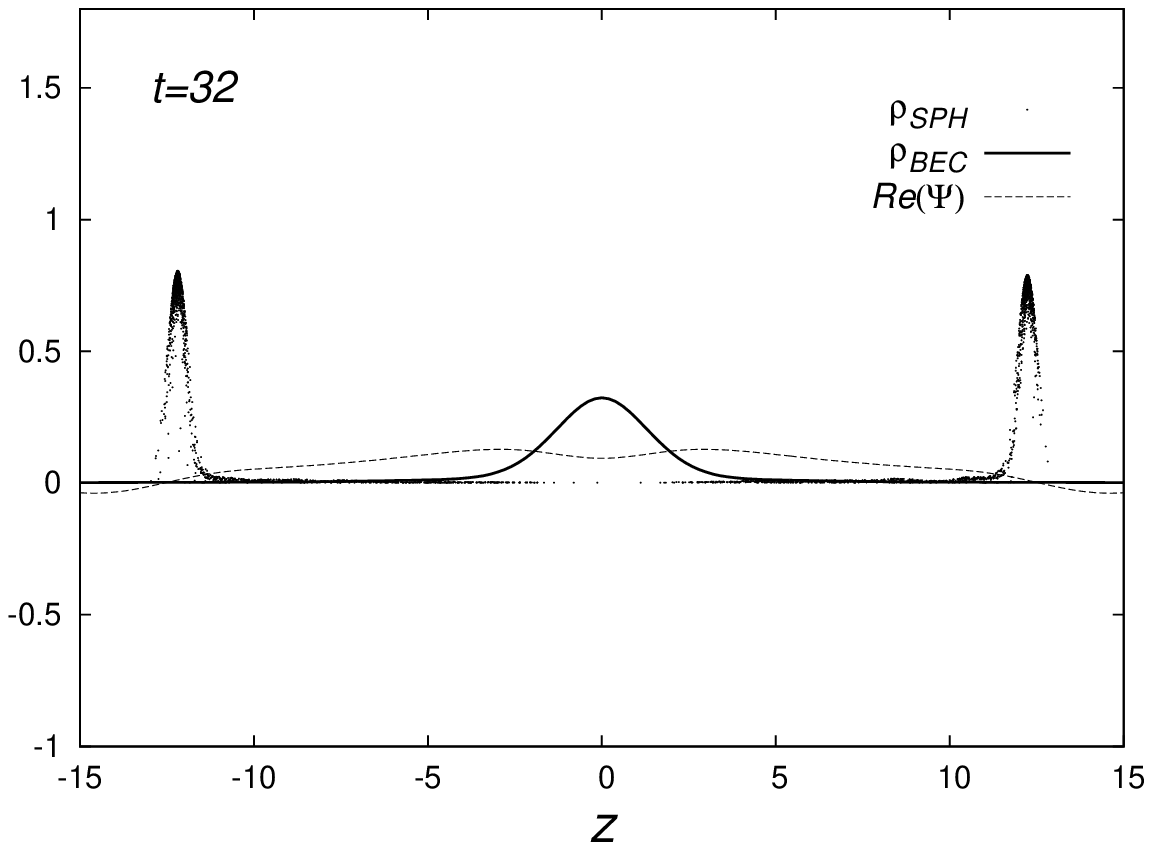}
\caption{\label{fig:collision2} We show snapshots of $\rho_{BEC}$, $\rho_{L}$ and $Re(\Psi)$ at various times for a collision with $v_z=0.5$ in code units. We use 5000 particles for the luminous matter.}
\end{figure}

As pointed out in  \cite{GonzalezGuzman2011}, it is interesting to measure the effects of the interference pattern shown by the BEC on the luminous matter. The main effect is that of the relative velocities of the luminous and BEC dark matter after the collision with solitonic behavior. The method we choose to evaluate the location of the luminous and dark components is the same described for the boosted configuration. We track the location of the maximum of the density profiles and determine how the BEC dark matter moves with respect to the luminous matter during and after the collision.

In order to systematize the analysis for various parameters of the encounters, we define $d_{A}$ and $d_B$ as the shift between the location of the BEC with respect to the luminous matter for the structure $A$ (the structure moving from left to right along the $z$ axis) and $B$ (the structure moving from right to left along the $z$ axis). Considering the structures have the same mass, then $d_A=d_B$, which reduces the problem to track $d_A$ only. In Fig. \ref{fig:d_A} we show $d_A$ for two cases. According to our definition, when $d_A>0$ the BEC travels in front of the luminous matter and when $d_A<0$ the BEC travels behind the luminous matter. The curves in Fig. \ref{fig:d_A} for $d_A$ show two branches, because we have removed the time domain in which the BEC shows the interference pattern and $\rho_{BEC}$ has a maximum at $z=0$. The line on the left shows $d_A$ prior to collision, showing a growth, which is already different from what was found in the test of a single boosted configuration, where $d_A \sim 0$. This means that the growth of $d_A$ is due to the influence of the potential wells of the two structures. The line on the right shows $d_A$ after the collision. For $v_z=1, ~1.5$, when $d_A>0$ the BEC moves in front of the luminous matter toward the right, as illustrated in Fig. \ref{fig:collision} at $t=11.2$ and $t=11.6$. However, when $d_A<0$ the luminous matter catches up with the BEC maximum and takes it over from then on as illustrated in Fig. \ref{fig:collision} at $t \sim 17.6$.

The interesting branch of $d_A$ to analyze is the one after the two BEC potential wells have passed through each other. In Fig. \ref{fig:d_A} can be seen that after the collision $d_A$ shows a linear behavior in time that we fit in order to model the progress in time of $d_A$ as a function of $v_z$ with a line $f(x)=px+q$. We thus show the branch of $d_A$ after the collision and fit with a line for various values of $v_z>0.755$ in Fig. \ref{fig:fits}. Notice that the off-set between luminous and dark matter is slower (small $p$) when $v_z$ is higher.

In order to illustrate the behavior of luminous and dark matter in a case with $E<0$, we show in Fig. \ref{fig:collision2} the evolution for the case $v_z=0.5$. Initially the luminous matter approaches the collision following the BEC, however after $t\sim 16$, $\rho_{BEC}$ forms a single high density configuration that evolves later on. Finally, a great part of the luminous matter escapes from the potential well and leaves a trace of disrupted matter near the potential well, indicated by the dots of $\rho_{L}$ dispersed over the domain.

\begin{figure}
\includegraphics[width=4.25cm]{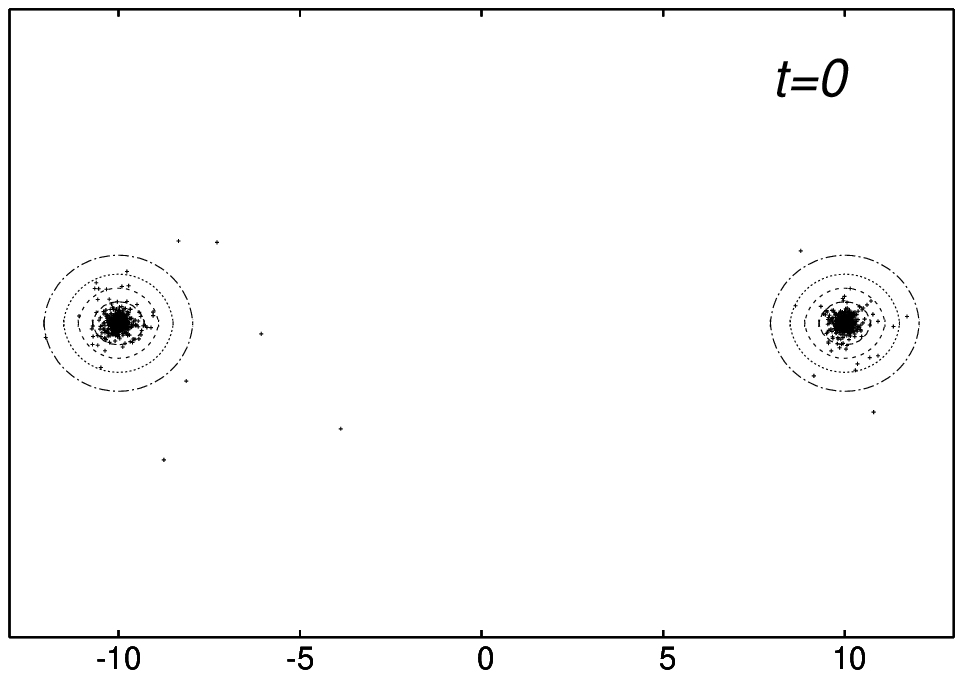}
\includegraphics[width=4.25cm]{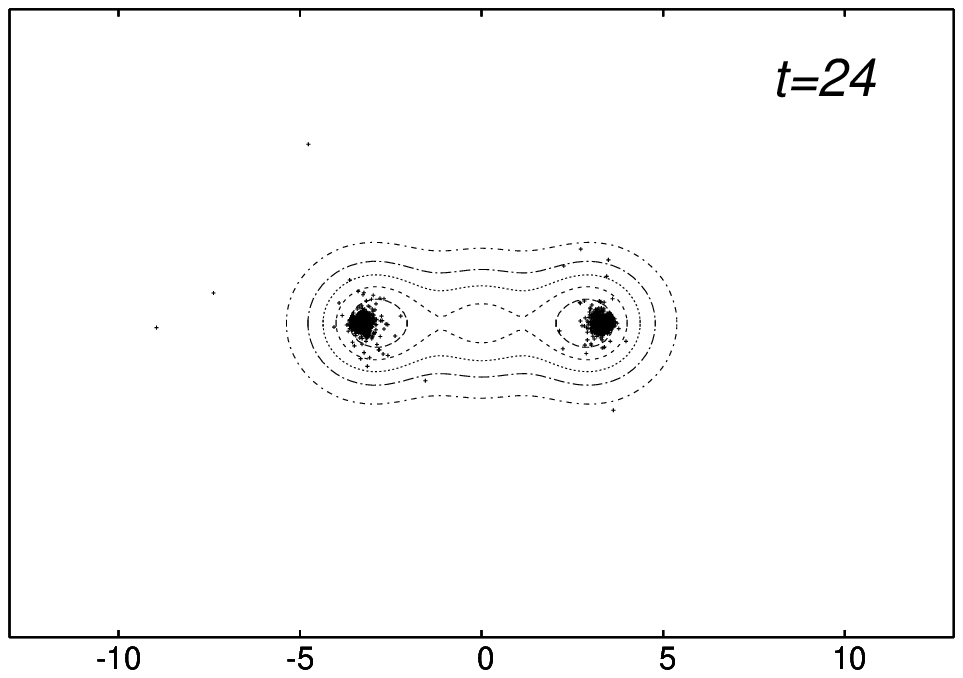}
\includegraphics[width=4.25cm]{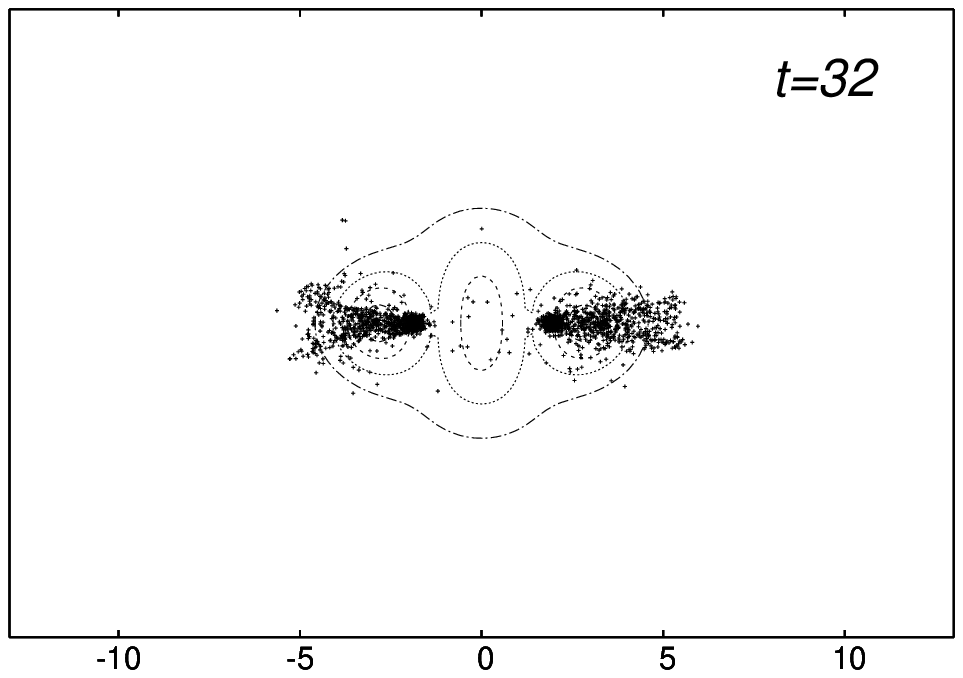}
\includegraphics[width=4.25cm]{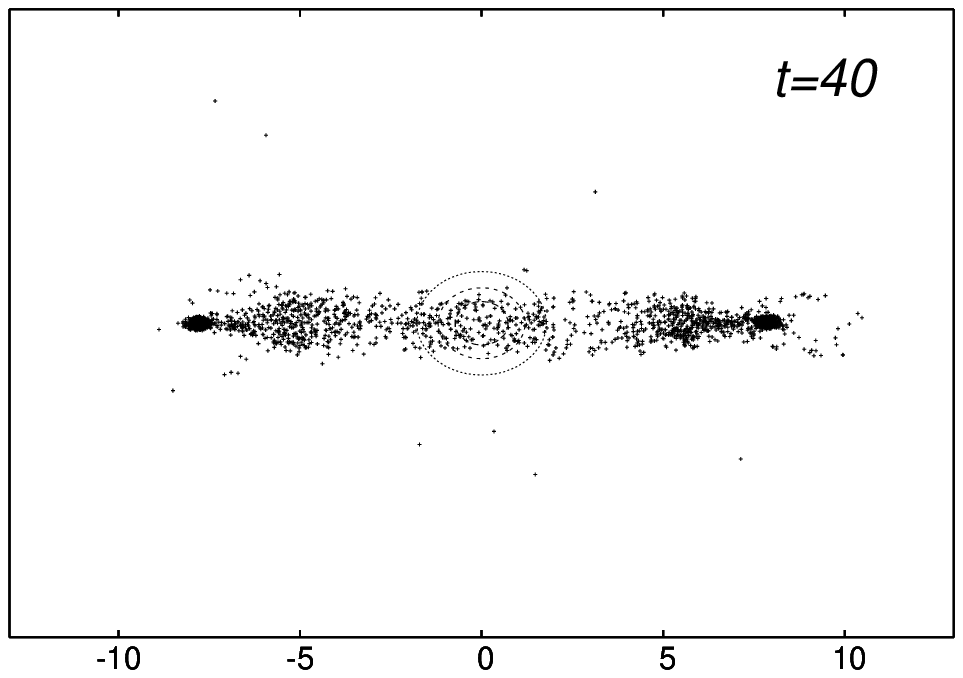}
\includegraphics[width=4.25cm]{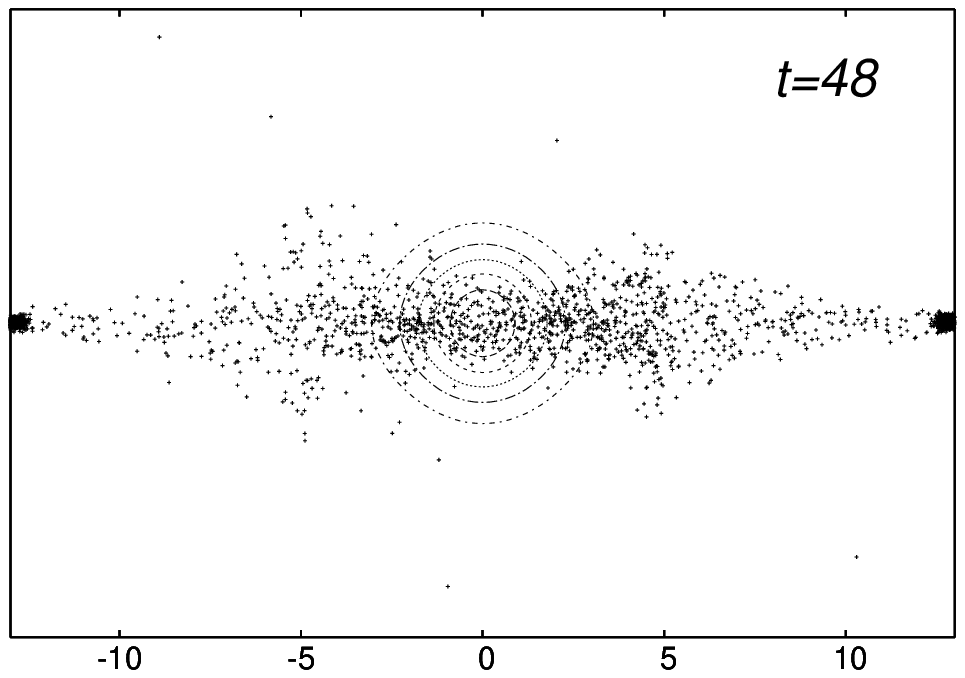}
\includegraphics[width=4.25cm]{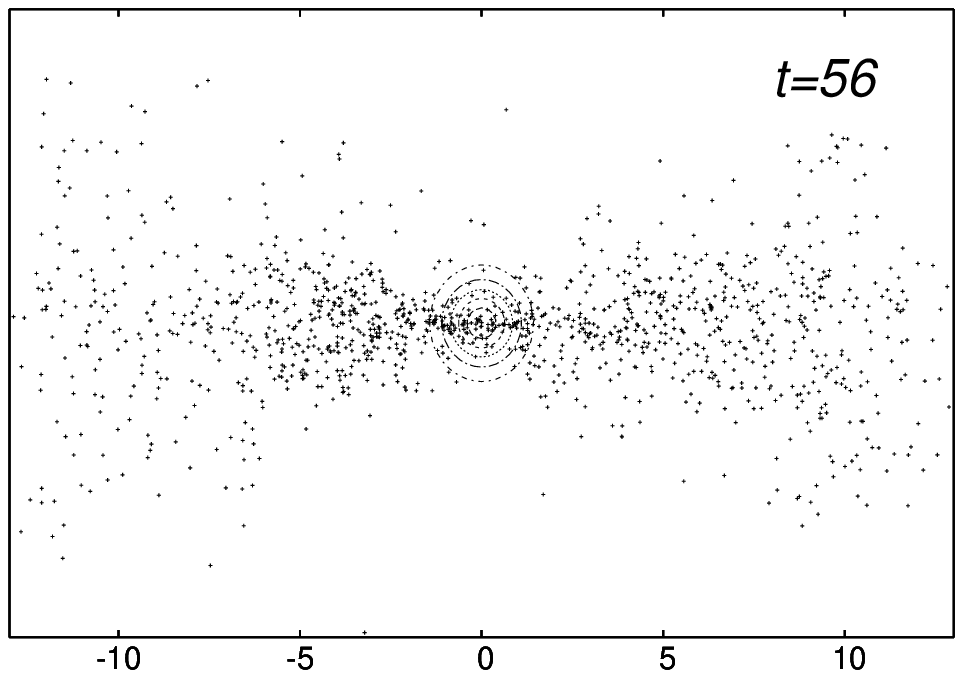}
\caption{\label{fig:physicalcase} Snapshots of the evolution of the Luminous matter and contours of $\rho_{BEC}$ that show their relative location for $v_z=0.2$. As the dark matter gets glued during the collision, the Luminous matter dilutes.}
\end{figure}

\begin{figure}
\includegraphics[width=8cm]{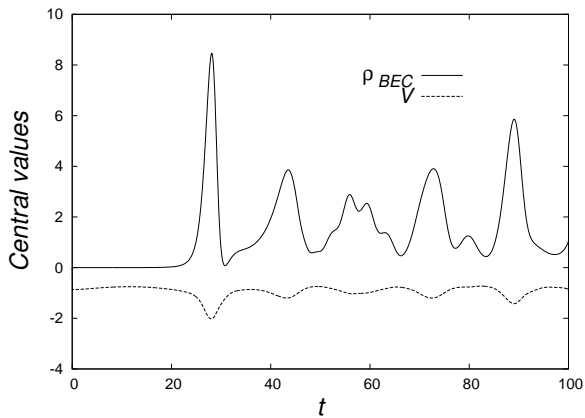}
\caption{\label{fig:Centrals} We show the central values of $\rho_{BEC}$ and $V$ in time, in code units, for the simulation in Fig. \ref{fig:physicalcase}.}
\end{figure}

\subsection{Particular cases}

{\it Case of solitonic behavior.} As mentioned before, the boson mass fixes the scales of the problem and a particular popular value of the boson mass is $m=10^{-23}eV/c^2$. Now, assuming for example that each structure has dark matter mass $\tilde{M}_{BEC}=10^{11}M_{\odot}$, using the scaling relations in \ref{subsec:GPP} and recovering physical units, we have that the initial velocity in our simulations, $\tilde{v}_z \simeq 1087.6 v_z [km/s]$. The solitonic behavior is expected to happen for $v_z>0.755$, that is, $\tilde{v}_{rel} > 1087.6 (0.755) km/s=821.14 km/s$, which actually would imply a relative minimum velocity of 1642$km/s$, a rather unexpectedly high velocity in dwarf galaxy collisions.

{\it Slow velocity collision with a single final halo}. To follow with the parameters of the previous example,  assume again that the boson mass is $m=10^{-23}eV/c^2$ and that each structure has dark matter mass $\tilde{M}_{BEC}=10^{11}M_{\odot}$ and $\tilde{v}_z \simeq 1087.6 v_z [km/s]$. Considering the case $v_z=0.2$, the initial head-on velocity of each structure is $\tilde{v}_z \simeq 217.52 [km/s]$. This is a very similar case to one presented in  \cite{Paredes2015} with the only difference that in that case the velocity is set to $\tilde{v}_z=200km/s$.

The evolution of this case is shown in Fig. \ref{fig:physicalcase}. To our surprise, the luminous matter concentrations remain for a little while after the collision and then they get dispersed away after the collision. The reason for the dispersion of luminous particles may be the incredible dynamics of the BEC. This is illustrated by the central value of $\rho_{BEC}$ and $V$ in Fig. \ref{fig:Centrals}. Prior to $t=20$ the encounter has not happened, and there the density at the center of mass is nearly zero, however by $t\sim 25$ the collision takes place. The oscillations of the central values start happening after the two blobs have merged and are glued together, and the single final configuration evolves by itself toward a relaxed state. The only studied relaxation mechanism for this system is the Gravitational Cooling \cite{SeidelSuen1990,GuzmanUrena2006}, which consists in the emission of $|\Psi|^2$ toward infinity, and in this case appears to be so violent that the luminous matter is shaken and expelled from the potential well.


\section{Final comments and conclusions}
\label{sec:conclusions}

We presented a study of head-on encounters of two equal mass BEC dark matter structures involving luminous matter under various dynamical conditions.

We analyzed the motion of luminous matter motion after the collision in two regimes. First in the case in which there is solitonic behavior, we found that after the encounter, the structures retain their basic structure of dark and luminous matter, like if they were nearly solitons, and an off-set in their relative motion is found as in \cite{Paredes2015}. The BEC first moves faster than the Luminous matter, and eventually the luminous matter takes over the dark matter potential wells.

Second, in the case of slow velocity encounters resulting in a single BEC configuration, we found that the luminous matter disperses away after the collision. This would be a fingerprint of BEC dark matter that deserves a more detailed analysis in specific observational cases. The explanation we find is the extremely restless behavior of the gravitational potential of the BEC component. Simulations of encounters of galaxies assuming CDM halos, show the luminous matter is not dispersed away (see e.g. \cite{nene}). The main difference with particle-like dark matter and BEC dark matter is essential, point-like particles interact and relax through gravitational interaction without any particular unexpected density pattern, whereas BEC dark matter shows interference patterns \cite{WaveDM} and uses the gravitational cooling as the only relaxation channel, which is a rather unquantified process whose effects have never been studied on particles around.
  
Our results were based on simulations carried out with a code adequately tested which is ready to tackle more realistic scenarios, including less aggressive encounters involving an impact parameter that should bring interesting results on the type of behavior of luminous matter in an encounter. Also, equal mass head-on encounters seem rather unfrequent and our scenario will be generalized. Another key issue that calls the attention is that the resulting structure after a collision is highly dynamical, and an estimate of time scale for relaxation would impact on the viability of the BEC dark matter model.

Before presenting a definitive answer it is important to enrich this model in various directions, namely, by adding the back reaction of luminous matter onto the gravitational potential of the BEC that should allow the study of encounters with more luminous matter and the influence of a non-zero self-interaction $a$ in encounters.


\section*{Acknowledgments}

This research is partly supported by grants CIC-UMSNH-4.9 and CIC-UMSNH-4.23.


\end{document}